\documentclass[aps,prd,twocolumn,showpacs,floatfix]{revtex4}
\usepackage{amsmath,amssymb}
\usepackage{bbm}
\usepackage{color}
\usepackage{graphicx}
\usepackage{times}
\newcommand{\ft}[2]{{\textstyle\frac{#1}{#2}}}
\newcommand{\half}{\frac{1}{2}}

\DeclareMathAlphabet{\boldmathe}{T1}{cmr}{bx}{it}
\DeclareMathAlphabet{\mathpzc}{OT1}{pzc}{m}{it}
\DeclareMathOperator{\tr}{tr}
\DeclareMathOperator{\Tr}{Tr}
\newcommand{\abs}[1]{\left| #1 \right|}

\newcommand{\mbf}[1]{\boldmathe{#1}}
\newlength{\graphwidth}
\newcommand{\refs}[1]{(\ref{#1})}
\newcommand{\eqnn}[1]{
\begin{equation}#1\end{equation}}
\newcommand{\eqnl}[2]{\begin{equation}#1
\label{#2}\end{equation}}
\newcommand{\eqngrl}[3]{\begin{equation}\begin{split}#1 \\ #2\end{split}\label{#3}\end{equation}}
\begin{document}
\def\pa{\partial}
\def\gA{^{g}\hskip-.7mm A}
\def\nA{\,^{n}\hskip-.6mm A}
\def\Azk{{^{z_k}}\hskip-.7mm A}
\def\Az{{^z}\!A}
\def\alphaA{\,^{\alpha}\hskip-.6mm A}
\def\hA{\,^{h}\hskip-.6mm A}
\def\gF{^{g}\hskip-.4mm F}
\def\gL{^{g}\hskip-.4mm L}
\def\gP{^{g}\hskip-.2mm {\cal P}}
\def\gpsi{^{g}\hskip-.2mm\psi}
\def\psiz{{^z}\psi}
\def\psizk{{^{z_k}\psi}}
\def\lamz{{^{z\!}\lambda}}
\def\lamzk{{^{z_k\!}\lambda}}
\def\rhoz{{^{z}\!\varrho}}
\def\sigz{{^{z}\!\varsigma}}
\def\rhozk{{^{z_k}\!\varrho}}
\def\rhobzk{{^{z_k}\!\bar\varrho}}
\def\diz{{^{z}\mathcal D}}
\def\dizk{{^{z_k}\mathcal D}}
\def\cC{{\mathcal C}}
\def\cD{{\mathcal D}}
\def\cDU{\cD_{U}}
\def\gcD{^{g}\hskip-.1mm \cD}
\def\ncD{^{n}\hskip-.1mm \cD}
\def\alphacD{^{\alpha}\hskip-.1mm \cD}
\def\hcD{^{h}\hskip-.1mm \cD}
\def\cA{{\mathcal A}}
\def\cE{{\mathcal E}}
\def\cF{{\mathcal F}}
\def\cG{{\mathcal G}}
\def\cP{{\mathcal P}}
\def\cQ{{\mathcal Q}}
\def\cS{{\mathcal S}}
\def\cW{{\mathcal W}}
\def\cZ{{\mathcal Z}}

\def\Z{\mathbb{Z}}
\def\T{\mathbb{T}}
\def\R{\mathbb{R}}

\def\id{\mathbbm{1}}
\def\mtxt#1{\quad\hbox{{#1}}\quad}
\def\mbe{\mbf{e}}
\def\mbx{\mbf{x}}
\def\mby{\mbf{y}}
\def\mbv{\mbf{v}}
\def\di{\slashed{D}}
\def\gdi{^{g}\hskip-.4mm\slashed{D}}
\def\lam{\lambda}

\title{Relation between chiral 
symmetry breaking and confinement in YM-theories }

\author{Franziska Synatschke,  Andreas Wipf}
\affiliation{Theoretisch-Physikalisches Institut,
Friedrich-Schiller-Universit{\"a}t Jena, Max-Wien-Platz 1, 07743
Jena, Germany}
\author{Kurt Langfeld}
\affiliation{School of Mathematics \& Statistics, University of Plymouth,
Plymouth, PL4 8AA, UK}
\begin{abstract}
Spectral sums of the Dirac-Wilson operator and their
relation to the Polyakov loop are thoroughly investigated. The 
approach by Gattringer is generalized to mode sums which reconstruct 
the Polyakov loop \textit{locally}. This opens the possibility to 
study the mode sum approximation to the Polyakov loop correlator. 
The approach is rederived for the ab initio continuum formulation of 
Yang-Mills theories, and the convergence of the mode sum is studied
in detail. The mode sums are then explicitly calculated for 
the Schwinger model and $SU(2)$ gauge theory in a homogeneous background 
field. Using $SU(2)$ lattice gauge theory, the IR dominated mode sums are 
considered and the mode sum approximation to the static quark antiquark 
potential is obtained numerically. We find a good agreement between the 
mode sum approximation and the static potential at large distances 
for the confinement and the high temperature plasma phase.
\end{abstract}
\pacs{12.38.Aw, 11.10.Wx, 11.15.Ha}
\maketitle

\section{Introduction}
Color confinement and spontaneous chiral symmetry breaking are
the two most relevant features of Yang-Mills theory when the
structure of matter under normal conditions is explored.
Both phenomenons are attributed to the low energy sector of Yang-Mills
theory, an analytic description of which is hardly feasible due to
strong couplings between the basic degrees of freedom, quarks and gluons.
Lattice gauge simulations found that both so different phenomenons
are intimately related: in the chiral limit, the critical temperature 
for deconfinement $T_d$ and the critical temperature for chiral restoration $T_c$
coincide~\cite{Karsch:1998ua}. This finding is highly nontrivial since the 
expectations had been that two totally different mechanisms involving 
different energy scales were at work for each phenomenon. Indeed, if 
quarks which transform under the adjoint representation were considered instead 
of the quark fields of standard QCD, largely different values for $T_d$ and $T_c$
were reported in~\cite{Karsch:1998qj}. 

These results have stirred the hope that a single low energy effective degree of freedom 
is responsible for both phenomenons. Given that in pure $SU(N)$ gauge theory 
the long distance part of the static quark potential depends on the $N$-ality of 
the quarks and that the crucial difference between fundamental and adjoint 
quarks is $N$-ality again, center vortices appear as a natural candidate 
for such a degree of freedom: these vortices are tightly related to confinement, 
are sensible in the continuum limit and offer an intriguing picture of deconfinement 
at high 
temperatures~\cite{DelDebbio:1996mh,Langfeld:1997jx,Greensite:2003bk,Engelhardt:1999fd}.
It was found quite recently that the vortices extend their 
reach to a description of spontaneous chiral symmetry breaking as 
well~\cite{Gattnar:2004gx}.

In order to reveal a model independent link between confinement and 
chiral symmetry breaking,  Gattringer proposed to reconstruct the Polyakov 
loop expectation value $\langle P \rangle $ in terms of a  particular spectral 
function $\cS_{N_t}$
of the lattice Dirac operator~\cite{Gattringer:2006ci}. 
While the low lying modes of the Dirac operator are directly related 
to chiral symmetry breaking by virtue of the Banks-Casher relation, 
the Polyakov loop expectation value serves as the litmus paper for confinement. 
Spectral representations of $\langle P \rangle $ have subsequently been 
the subject of recent studies. Bruckmann et al.~\cite{Bruckmann:2006kx}
investigated the response of the eigenvalues of the staggered Dirac
operator to a twist of the boundary conditions. They found that
the infrared part of the spectrum is most sensitive to twists. 
It was subsequently pointed out that the spectral sum $\cS_{N_t}$, originally 
proposed by Gattringer, is dominated by the ultraviolet part of the spectrum 
since the sum contains large powers of the 
eigenvalues~\cite{Gattringer:2006ci,Bruckmann:2006kx,Synatschke:2007bz,Soldner:2007vp}.
A sensible continuum limit was caught into question. 
In~\cite{Synatschke:2007bz} alternative spectral sums were put forward which 
serve as order parameters for confinement and which receive their
main contributions from the infrared part of the spectrum.
One of these sums is of particular interest since it relates
the dressed Polyakov loops to the chiral condensate via
the celebrated Banks-Casher relation \cite{Hagen:2007hz}.
If one twists the gauge field by a complex number
$z$ with unit modulus (or equivalently
twists the boundary conditions by $1/z$) and picks the
coefficient of $z^k$ in the spectral sums one
obtains the dressed Polyakov loops with winding number $k$
\cite{Synatschke:2007bz}. With
this method Bilgici et al. \cite{Bilgici:2008qy} connect the eigenvalue
density at zero and therefore the chiral condensate to the dressed Polyakov
loops. Numerical results for spectral sums of various
lattice Dirac operators with quenched configurations can
be found in the recent papers \cite{Bruckmann:2006kx,Synatschke:2007bz,Hagen:2007hz,Bruckmann:2008rj}
and for ensembles generated with dynamical fermions
in \cite{Soldner:2007vp}. 
We finally point out that the spectral approach to the Polyakov line 
has been extended  by using eigenmodes of the Laplacian 
operator~\cite{Bruckmann:2005hy}. 
There, the spectral sum acts as gauge invariant low energy filter 
which reveals the ``classical'' texture while quark confinement is 
still active. 

In this paper, we generalize the concept of the mode sum approximation 
to reconstruct the Polyakov loop \textit{locally}. This has the great advantage 
that now the Polyakov loop correlator and the static quark antiquark 
potential can be studied in the light of a few low lying modes of the Dirac
operator. We then point out that the mode sum approach is not solely tied to 
lattice quark operators and present an explicit construction of this approach 
in the ab initio continuum formulation of Yang-Mills theory. 
The convergence of the mode sums \emph{for all} polynomial functions of the
continuum Dirac operator is demonstrated for the first time. 
We then argue that a wide class of IR dominated mode sums are in fact 
proportional to the Polyakov loop. This conjecture is fostered by an explicit 
calculation of these mode sums for Schwinger model and for $SU(2)$ gauge theory 
with constant background field strength. 
We then consider $SU(2)$ gauge theory above and below the deconfinement 
temperature by means of lattice gauge simulations. Most important, we find 
that a few low lying modes of the quark operator are sufficient to 
reconstruct the static quark potential at large quark antiquark distances. 
This is a gauge invariant and model independent signal that the 
color confinement mechanism has its fingerprints in the low lying 
quark spectrum.

\section{Recall of spectral sums for Lattice models}
\begin{figure}[ht]
\includegraphics[width=8.6cm]{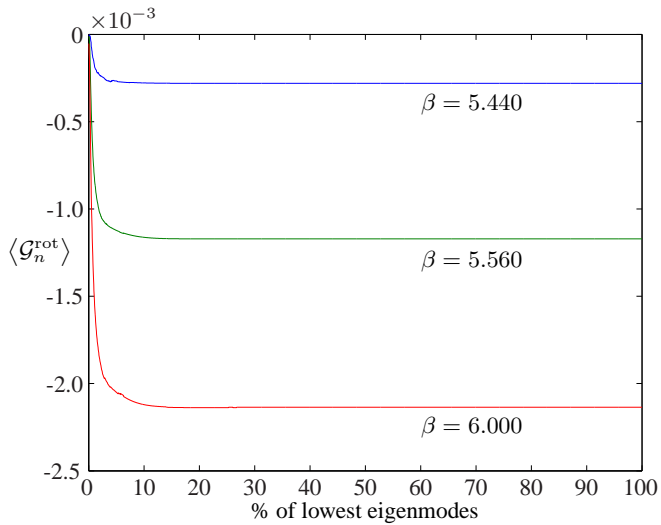}
\caption{\label{fig:expoMinusLaLaQuer43} 
Mean Gaussian sums $\cG^{\rm rot}_n$ for $SU(3)$ on a $4^3\!\times\!3$-lattice
near $\beta_{\rm crit}$. The graphs are labeled with $\beta$.}
\end{figure}

The $j$-th power of a Dirac operator $\cDU$
with nearest neighbor interaction on
an Euclidean lattice with $N_t\times N_s^{d-1}$ sites
can be expanded in Wilson loops of length up to $j$,
\eqnl{
\langle x\vert \cDU^{\,j}\vert x\rangle
=\sum_{\vert\cC_{x}\vert\leq j} a_{\cC_{x}}\cW_{\cC_{x}}.
}{lat1}
The value of the coefficient $a_{\cC_{x}}$
multiplying the holonomy $\cW_{\cC_x}$ of the loop $\cC_x$ with base $x$
depends on the type of fermions under consideration.
The expansion \refs{lat1} is used to relate the
Polyakov loop
\eqnl{
P(\mbx)=\tr \cP(\mbx),\quad \cP(\mbx)=\prod_{x_0}U_0(x_0,\mbx)}{lat3}
with the spectrum of the Dirac operator
\cite{Gattringer:2006ci}.
On a lattice with $N_s\gg N_t$ the sum on the right hand
side of \refs{lat1} contains
contractable loops and loops winding once or several
times around the torus in time direction. 
In order to relate the Polyakov loop to spectral sums of quark eigenmodes
an interface is inserted into the lattice gauge  configuration 
$U_\mu(x),\; x= (x_0,\mbx)$ by
\eqnl{
{^z}U_\mu (x_0,\mbx) =\left\{ \begin{array}{cl}
z\cdot U_0 (x_0,\mbx) & \hbox{ for $\mu =0 $ and $x_0=0$ } \\
U_\mu (x_0,\mbx) & \hbox{ otherwise }, \\
\end{array} \right. }{lat5}
where $z=e^{2\pi i\alpha}$ is a center element. Contractable
Wilson loops are invariant when one inserts an interface
which is referred to as \emph{twisting the gauge field.}
Wilson loops winding $k$-times acquire a factor
$z^k$. It follows that for a twisted gauge field the coefficient of $z$ 
in the series \refs{lat1} becomes a linear combination 
of dressed Polyakov loops passing through $x$,
having length $\leq j$ and  winding $1+kN$ times
around the time direction. Here $k$ is an integer and $z^N=1$. 
For $j=N_t$ there is only one such loop, namely the straight Polyakov
loop at  $\mbx$. After taking the traces over
spinor and color-indices the sum over $x^0$ yields
\cite{Gattringer:2006ci}
\begin{gather}
P(\mbx)=\frac{1}{\kappa}
\sum_{k=1}^N  z_k^\ast
\sum_{p=1}^{n_p} \rhozk_p(\mbx)\,\left(\lamzk_p\right)^{N_t}\label{LatSum1}
,\\ \rhoz_p(\mbx)=\sum_{x_0=1}^{N_t}\rhoz_p(x^0,\mbx).\label{LatSum1b}
\end{gather}
The first sum in~\refs{LatSum1} is over all center elements $z_1,\dots,z_N$ and
the second
over all $n_p$ eigenvalues of the Dirac operator.
The value of the constant $\kappa$ depends on the type of
lattice Dirac operator under consideration. $\rhoz_p(x)$ is the eigenvalue
density
and $\lamz_p$ the $p$-th eigenvalue of the Dirac
operator $^{z}\cDU\equiv \cD_{^{z}U}$ with $z$-twisted gauge field,
\begin{equation}
\begin{split}
(^{z}\cDU)\,&\psiz_p=\lamz_p\psiz_p \mtxt{with}\\
&\psiz_p(x_0+N_t,\mbx)=-\psiz_p(x_0,\mbx). \label{eig1}
\end{split}
\end{equation}
In terms of the normalized eigenmodes the color-blind density reads
\eqnl{
\rhoz_p(x)=\sum_{\ell}\left\vert\psiz_{p,\ell}(x)\right\vert^2,
}{dens1}
where the sum extends over all eigenfunctions of $^{z}\cDU$ with fixed energy
$\lam_p$. The densities $\rhoz_p$ are gauge invariant scalar fields.
For the trivial center element $z=1$ we often write $\,\varrho$
instead of $\rhoz$. 
Averaging the local identity \refs{LatSum1} over space yields
\eqnl{
\bar P\equiv\frac{1}{V_s}\sum_\mbx P(\mbx)=\frac{1}{\kappa}
\sum_k z_k^\ast
\sum_p \left(\lamzk_p\right)^{N_t} .}{LatSum2}
This simple formula for the averaged loop has been investigated
a lot in the past. The main problem with the sum on
the right hand side is that it is dominated by the ultraviolet part of the spectrum
and therefore is expected to have an ill defined continuum
limit. But  \emph{all} spectral sums of the form
\eqngrl{
\cS_f(U)&= \sum_k z_k^\ast \sum_p f(\lamzk_p)=\sum_x \cS_f(U;x)}
{\cS_f(U;x)&= \sum_{k}  z_k^\ast
\sum_{p=1}^{n_p} \rhozk_p(x)\,f(\lamzk_p)\,}{LatSum3}
define (nonlocal) order parameters for the center symmetry
\cite{Synatschke:2007bz}.
Indeed, if we twist the gauge field with a center element $z$, we obtain: 
\eqnl{
\cS_f(^zU)=z \cS_f(U).}{LatSum4}
Our important observation is that, as the Polyakov loop, 
\emph{all} spectral sums pick up a factor in the center of the group. 
Thus, not only the Polyakov loop, but also any other 
spectral sum of the above type might serve as a litmus paper for 
confinement. 
Of particular interest are sums which get their main
contribution from the low lying eigenvalues. It has
been convincingly demonstrated in \cite{Synatschke:2007bz}
that the Gaussian sum with $f(\cD)=\exp(-\cD\cD^{\,\dagger})$ is very
well suited for that purpose. For a $SU(3)$ lattice gauge theory, 
Figure \ref{fig:expoMinusLaLaQuer43}
shows the Monte-Carlo averages of the \emph{partial sums}
\eqnl{
\cG_n(U)=\sum_{k=1}^3  z_k^\ast \sum_{p=1}^n
\exp\left(-\vert \lamzk_p\vert^2\right)
}{LatSum5}
and demonstrates their rapid convergence to a multiple of the 
(rotated) Polyakov loop expectation value. Actually it is sufficient 
to include less than $5\%$ of the low lying 
eigenvalues to obtain a decent approximation of  the limiting value.
Similar results hold for the spectral sums of the
functions $f(\cD)=\cD^{\,-1}$ and $f(\cD)=\cD^{\,-2}$
corresponding to the propagators of $\cD$ and $\cD^2$.
They are of particular interest since they relate to
the celebrated Banks-Casher relation.

The spectral problem \refs{eig1} is gauge-equivalent to the problem
with twisted boundary conditions
\begin{equation}
\begin{split}
\cDU&\psiz_p=\lamz_p\psiz_p\mtxt{with}\\
&\psiz_p(x_0+N_t,\mbx)=- z^\ast\,\psiz_p(x_0,\mbx).
\end{split}
\label{eig3}
\end{equation}
For calculating the gauge invariant spectral sums
we may either twist the gauge field as in \refs{lat5}
or twist the boundary conditions as in
\refs{eig3}. Bilgici et al. extended the results in \cite{Synatschke:2007bz}
and allowed for twists of the type $z=\exp(2\pi i\alpha)$ in the  
boundary conditions \cite{Bilgici:2008qy}. Admitting arbitrary values 
$\alpha \in [0, 1[$,  the twists are no longer center elements of $SU(N)$
but only of $U(N)$.
Using nevertheless (\ref{eig3}) as the primary definition 
of twist when $z$ is extended to $U(1)$ phases, the coefficients 
$\tilde\Sigma_n$ of the expansion of the twisted quark propagator in 
powers of $z$ can be easily obtained \cite{Bilgici:2008qy}. 
The coefficient $\tilde \Sigma_n$ can be written as sums over all loops
winding $n$-times around the torus in time direction. 
In particular, the coefficient $\tilde \Sigma_1$ is related to the spacetime
integrated 
spectral density $\rhoz (\lambda ) $ for vanishing eigenvalue by 
\eqnl{
\tilde\Sigma_1=\int_0^{1} d\alpha\;
z^\ast\cdot\rhoz (\lam\searrow 0),\quad z=e^{2\pi i\alpha}.}{LatSum6}

In the numerical investigation below we will
focus on truncations of \textit{local} spectral sums 
(\ref{LatSum3}). We will use IR sensitive spectral sums to find good 
approximations to $\langle P(\mbx)P(\mby)\rangle$ or higher
correlators. Guided by our previous results we consider the 
\emph{partial} Gaussian sums
\eqnl{
\cG_n(U;\mbx)=\sum_k z^\ast_k
\sum_{p=1}^n\!\rhozk_p(\mbx)\;
\exp\left(-\vert \lamzk_p\vert^2\right),}{LatSum8}
where $\varrho_p(\mbx)$ has been introduced in \refs{LatSum1b} and \refs{dens1}.
We will study by analytical (for cases which admit a complete analytical 
evaluation of the spectral sums) and by numerical means 
whether and, in case, to which extent the correlator of two partial sums 
approaches the Polyakov loop correlator 
\eqnl{
\langle \cG_n(U;\mbx)\cG_n(U;\mby)\rangle
\longrightarrow
\hbox{const}\cdot\langle P(\mbx)P(\mby)\rangle,
}{LatSum9}
when $n$ tends to the total number $n_p$ of eigenvalues. 
\section{Spectral Sums and Center Symmetry in the Continuum}
So far the intriguing relations between spectral sums of twisted
configurations and Polyakov loops have been established
for lattice regulated gauge theories only. 
It is yet an open question 
which results remain meaningful in the continuum limit.
Clearly, an object like $\Tr (\cD_U^{N_t})$ does not make sense
in the continuum, and this was the main motivation 
in  \cite{Synatschke:2007bz} to introduce the generalized spectral 
sums  \refs{LatSum3}. Even if $\,\Tr f(\cD)$ exists
in the continuum limit and even if, on the lattice,  
\eqnl{
\cS_f(U;\mbx)=\sum_{x^0} \cS_f(x^0,\mbx;U)}{intro0}
with $\cS_f(x;U)$ from \refs{LatSum3}
is roughly proportional to the Polyakov loop, there are still 
the possibilities that 
this \emph{approximate} proportionality is lost in the 
continuum limit or that the constant of proportionality 
diverges. For example, it was observed that the factor $\kappa$ 
in \refs{LatSum1} diverges in the continuum limit. 

In this section we study spectral sums for Euclidean gauge theories 
in the ab initio \emph{continuum} formulation for a torus 
of extend $\beta\times L^{d-1}$ with $L$ much bigger than the
inverse temperature $\beta=1/k_BT$. The
volume of the torus is $V=\beta\cdot V_s$ while the  spatial
volume is given by $V_s=L^{d-1}$. On a torus the continuum Dirac operator
\eqnl{
\cD_A=i\gamma^\mu(\partial_\mu-iA_\mu)+im}{intro1}
has discrete eigenvalues $\lam_p$ which are real
for a vanishing quark mass. We shall consider Hermitian vector potentials
\eqnl{
A_\mu=A^a_\mu \lam^a.}{intro3}
The gauge field is described by real-valued
functions $A^a_\mu(x)$ (with $a\in\{1,2,\dots,\dim(G)\}$)
of the Euclidean space-time points $x=(x_0,\mbx)$. The
path integral measure contains fields that obey periodic
boundary conditions in the Euclidean time direction,
\eqnl{
A_\mu(x_0+\beta,\mbx)=A_\mu(x_0,\mbx).}{intro5}
Even for configurations with nonvanishing instanton
number we may assume periodicity in the time direction
\cite{vanBaal:1982ag,Ford:1998bt}.
The Yang-Mills action is invariant under gauge transformations
\eqnl{
\gA_\mu=g(A_\mu+i\pa_\mu)g^{-1},\quad g(x)\in G,}{intro7}
under which the field strength transforms as
$\gF_{\mu\nu}=gF_{\mu\nu}g^{-1}.$
In order to maintain the boundary condition \refs{intro5} for the vector
potential, we must demand that the gauge transformations are
periodic up to a constant twist matrix $z$,
\eqnl{
g(x_0+\beta,\mbx)=z g(x_0,\mbx).}{intro11}
When such a topologically nontrivial
transformation is applied to a strictly periodic
vector potential $A_\mu$ then
\eqnl{
\gA_\mu(x_0+\beta,\mbx)=z\, \gA_\mu(x_0,\mbx) z^{-1}.}{intro13}
The gauge transformed potentials obey the boundary condition
\refs{intro5} only if $z$ commutes with $\gA_\mu$.
This limits us to twist matrices in the center of the gauge group
and explains why twisted gauge transformations are called
center transformations. In an irreducible and unitary
representation of the group
a center element is a multiple of the identity, $z=\hbox{phase factor}\cdot\id$.
The phase factor is such that $z$ is a group element. We shall
denote both the center element and its phase factor by $z$.

In the absence of matter fields the twisted transformations
form the global center symmetry. It can break spontaneously
and the traced Polyakov loop
\eqnl{
P(\mbx)=\tr \cP(\mbx),\quad
\cP(\mbx)=\cP \exp\left(i\int_0^\beta d\tau A_0(\tau,\mbx)\right),}{intro17}
which transforms nontrivially under center transformations
\eqnl{
{^g}\hskip-.4mm P(\mbx)=\tr \cP \exp\left(i\int_0^\beta d s\; \gA_0(s,\mbx)\right)
=z\, P(\mbx)
}{intro19}
serves as an order parameter for the center symmetry.
As is well-known, the expectation value of $P(\mbx)$ is nonzero
in the deconfining high-temperature phase and it is zero in
the confining low-temperature phase.

It is important to note that the center symmetry is explicitly
broken in the presence of matter fields in the fundamental
representation. For example, quark fields transform as
\eqnl{
\gpsi(x)=g(x)\psi(x)}{einf1}
and are antiperiodic in Euclidean time.
For a nontrivial twist the transformed field is not antiperiodic anymore,
\eqnl{
\gpsi(x_0+\beta,\mbx)=-z^{-1}\,\gpsi(x_0,\mbx).}{einf5}
The eigenvalues of the Dirac operator for the transformed gauge field
\eqnl{
\gcD_A \equiv \cD_{\gA}=g\cD_A g^{-1}}{einf7}
are different to the eigenvalues of $\cD_A$ if $g$ is
nonperiodic in time. Although
\eqnl{\cD_A\psi_p=\lam_p\psi_p}{einf9}
implies ${\gcD}_A\,{\gpsi_p}=\lam_p {\gpsi_p}$,
for a nontrivial twist the $\gpsi_p$ are no eigenmodes of ${\gcD}_A$ because
they are not antiperiodic in time.
But for two gauge transformations
$g$ and $\tilde g$ with the same twist $z$ in \refs{intro11}
 the transformed operators do have identical eigenvalues
\eqnl{
\lam_p(\gcD_A)=\lam_p({^{\tilde g}}\cD_A),}{einf11}
since the Dirac operators are gauge-related by the periodic gauge
transformation $\tilde g g^{-1}$.

Following the suggestion in \cite{Synatschke:2007bz}
we consider the weighted sums
\begin{equation}
\begin{split}
\cS_f(A) &=\sum_k z^\ast_k\,
\Tr f\left(\dizk_A\right)=\int d^dx\, \cS_f(A;x)\\
\cS_f(A;x)&=
\sum_k z^\ast_k\, \big\langle x\vert  
\tr f\left(\dizk_A\right)\vert x\big\rangle\\
&=\sum_k z^\ast_k \sum_{p=0}^\infty\rhozk_p(x)\,f(\lamzk_p)
,
\end{split}
\label{sun11}
\end{equation}
but now for continuum Dirac operators.
Here $\Tr$ denotes the trace over all
degrees of freedom,
whereas $\tr$ denotes the trace in spinor- and color space
only. Similarly as on the lattice one collects
the contribution to the spectral density of 
all eigenfunction with the same energy,
\eqnl{
\rhoz_p(x)=\sum_{\ell}\left\vert\hskip.2mm\psiz_{p,\ell}(x)\right\vert^2.}{sun12}
The $\rhoz$ are gauge invariant scalars which
transform nontrivially under center transformations.
According to a theorem of H. Weyl \cite{Weyl} the eigenvalues
of $\cD_A$ on a space of finite volume have the asymptotic distribution
$\lam_p\sim p^{-1/d}$
such that the traces in \refs{sun11} exist for functions $f$ which
decay faster than $1/\lam^{d}$ for large $\lambda$. Actually,
later we shall prove that the spectral sums defined as
\eqnl{
\cS_f(A;x)=\lim_{n\to \infty} \sum_{p=0}^n
\sum_k z^\ast_k\rhozk_p(x)\,f(\lamzk_p)}{sun13}
exist for a much bigger class of functions.

The spectral  function $\cS_f$ transforms under center transformations 
as follows,
\begin{equation}
\begin{split}
\cS_f(\gA)&=
\sum_k z^\ast_k\, \Tr f\left(\dizk_{\gA}\right)\\&=
\sum_k z^\ast_k\, \Tr f\left(^{zz_k}\cD_A\right)\\
&=
z\sum_k {z^\ast_k}^{\,\prime}\, \Tr f\left(^{z'_k}\cD_A\right)=
z \cS_f(A),
\end{split}
\label{sun15}
\end{equation}
where we have set $zz_k=z'_k$.  The same argument applies
to the density such that for all elements of the center
we have
\eqnl{
\cS_f(\gA)=z\cdot \cS_f(A)\mtxt{and}
\cS_f(\gA;x)=z\cdot \cS_f(A;x).
}{sun17}
All spectral sums $\cS_f$ transform the
same way as the Polyakov loop under center transformations 
and thus equally well serve as \emph{order parameters} for the center 
symmetry.
As on the lattice the eigenvalue problem for  $\,\gcD_A$ acting on antiperiodic 
functions is gauge-equivalent to
\eqnl{
\cD_A\psi_p=\lam_p\psi_p ,\quad\psi(x_0+\beta,\mbx)=-z^{-1}\psi(x_0,\mbx).}
{einf13}
For spectral problems the twisting of $A_\mu$ has the same
effect as twisting the boundary conditions with the inverse center 
element.
Since the shifts $\lam_p(\gcD_A)-\lam_p(\cD_A)$
only depend on the twist $z$ we may as well choose a simple
representative in every class of gauge transformations
characterized by this twist. 
\paragraph*{Gauge group SU(N):}
The cyclic center $\Z_N$ of this group is generated by
\eqnl{
z=\exp\big(2\pi i T/N\big)
\mtxt{with}
T=\hbox{diag}\,(1,1,\dots,1,1-N).
}{sun3}
As simple gauge transformations with twist $z_k=z^k$
we choose the powers $h^k$ of
\eqnl{
h(x_0)=\exp\big(2\pi i x_0 T/\beta N\big).
}{sun5}
The transformed gauge potential reads
\begin{multline}
\Azk_\mu=h^k(x_0) A_\mu(x) h^{-k}(x_0)+
\frac{2\pi k}{\beta N}\,T\delta_{\mu,0},\\ k=1,\dots,N,
\label{sun7}
\end{multline}
and the corresponding twisted Dirac operators are
\eqnl{
\dizk_A\equiv h^k \cD_A h^{-k}.
}{sun9}
\paragraph*{Gauge group U(N):}
The center of this group consists of the elements 
$z=\hbox{phase factor}\cdot\id$ with
arbitrary phase factors $e^{2\pi i\alpha}$. As simple
representatives for the twisted gauge transformations with
twist $z$ we choose
\eqnl{
h^\alpha(x_0)=\exp(2\pi i \alpha x_0/\beta)\cdot\id,\quad 0\leq \alpha\leq 1.
}{un1}
The gauge transformation shifts the potential by a constant
proportional to the identity matrix,
\eqnl{
\Az_\mu(x)=
A_\mu(x)+
\frac{2\pi}{\beta}\alpha\id\, \delta_{\mu,0},}{un3}
similarly as in the construction of the Nahm-transform
of selfdual $U(N)$-gauge fields \cite{Nahm:1979yw}.
The twisted Dirac operators are
\eqnl{
\diz_A\equiv h^\alpha \cD_A h^{-\alpha},}
{un5}
and the sum over the center elements in
the spectral function \refs{sun11} turns into an integral
\eqngrl{
\cS_f(A)&= \int d^dx\,\cS_f(A;x),\mtxt{with}}
{\cS_f(A;x) &=\int_0^1 d\alpha\, z^\ast\,
\langle x\vert\, \tr f\big(\diz_A\big)\vert x\rangle,\quad
z=e^{2\pi i\alpha}.}{un7}
Although the center is not discrete the transformation rule
for the spectral sums \refs{sun17} applies.
\paragraph*{Comparing SU(N) and U(N):}
The equivalence of the spectral problems for $\diz_A$ on
antiperiodic functions and $\cD_A$ on  functions with 
twisted boundary conditions \refs{einf13} can be used to prove 
that the Dirac operators for certain $su(N)$ and $u(N)$ 
fields have identical spectra. To show this
we consider a \emph{traceless} potential $A_\mu$ which can be
viewed both as $su(N)$ or as $u(N)$ potential. We transform
it with twisted gauge transformations $g\in SU(N)$
and $\tilde g\in U(N)$. The transformed potentials
$\gA_\mu\in su(N)$ and $\,^{\tilde g}\!A_\mu\in u(N)$ are
in general different. However, if $g$ and $\tilde g$ are twisted with
the same center element $z$ of $SU(N)$ then
\eqnl{
\lam_p(\gcD_A)=\lam_p(^{\tilde g}\cD_A).}{einf15}
Note that for $\alpha=k/N$ the
center element $h^\alpha(\beta)$ in \refs{un1} is actually in $SU(N)$ 
and the result \refs{einf15} applies. Thus for
any $su(N)$-potential $A_\mu$ the Dirac operators
with transformed potentials
\begin{equation}
\begin{split}
A^{(1)}_\mu(x)=&
e^{2\pi i k x_0T/\beta N} A_\mu(x)\, e^{-2\pi ikx_0T/\beta N}\\&+
\frac{2\pi k}{\beta N}\,T\delta_{\mu,0}\in su(N)\\
A^{(2)}_\mu(x)=&A_\mu(x)+
\frac{2\pi k}{\beta N}\id\, \delta_{\mu,0}\in u(N)
\end{split}\label{un9}
\end{equation}
have identical spectra. This observation is useful when
one calculates spectral sums.
\subsection{Spectral sums and Polyakov Loop}
In the absence of matter a gauge invariant function is
a function of the gauge invariant Wilson loops based 
at some fixed base point,
\eqnl{
W_{\cC_x}(A)=\tr\, \cP \left(\exp i\int_{\cC_x} A_\mu dx^\mu\right).
}{sl1}
For a contractable loop these objects are invariant
under center transformations and for loops winding
$k$-times around the periodic time directions they
pick up the factor $z^k$. We assume $L\gg \beta$ in 
which case we may neglect Wilson loops winding around 
the spatial directions. 

Since $\cS_f(A;x)$ is gauge invariant and transforms 
under center transformation the same way as a
dressed $P_{\cC_\mbx}$ with
base point $x=(0,\mbx)$, we conclude immediately
that the functions  $F(\cC_\mbx,A)$ in the expansion
\eqnl{
\cS_f(A;\mbx)=\sum_{\cC_\mbx} P_{\cC_\mbx}\cdot F(\cC_\mbx,A)
}{sl5}
are invariant under both twisted and periodic gauge transformations.
We conclude that for $SU(N)$ these functions
only depend on center symmetric Wilson loops based at $(0,\mbx)$.
For example, they may still depend on Wilson loops
winding $N$-times around the periodic time direction.
For the group $U(N)$ with continuous center $U(1)$ the function can only
depend on contractable Wilson loops and center symmetric
combinations $P^*_{\cC_\mbx}P_{\cC'_\mbx}$.

Actually, at least for the instanton solutions on the torus
constructed by 't Hooft \cite{'tHooft:1981ht}, a stronger result
holds true, namely
\eqnl{
\cS_f(A;\mbx)\stackrel{L\gg\beta}{\longrightarrow}\hbox{const}\cdot P(\mbx),
}{sl7}
and this will be demonstrated in the following section. Since a similar
relation approximately holds for certain spectral sums on the lattice one 
may conjecture that it holds for suitable chosen spectral sums in the 
continuum as well.

\subsection{On the convergence of spectral sums}
To investigate the convergence of the spectral sum we consider
the Gaussian spectral sum
\eqngrl{
\cG'(t,A) &=\int d^d x\, \cG'(t,A;x)}
{\cG'(t,A;x)&=
\sum_k z^\ast_k  \big\langle x\vert  \tr \exp\left(-t\,\dizk^2_A\right)\vert x\big\rangle
-\bar \varrho_0(x).}{conv1}
We subtracted the  center-averaged density $\bar\varrho_0(x)$ of the zero-modes
for later use. More generally, if $\psi_{p,\ell}(x)$ are the orthonormal 
eigenfunctions of $\cD_A^2$  with eigenvalue $\mu_p=\lambda_p^2$, 
then the center-averaged densities $\bar\varrho_p$ are
\eqnl{
\bar\varrho_p(x) =\sum_k z^\ast_k\;^{z_k}
 \hskip-.5mm \varrho_p(x)
}{conv11}
with $\rhoz_p(x)$ defined 
in (\ref{sun12}), wherein the $\psiz_{p,\ell}(x)$ are eigenfunctions
of $\diz_A^2$ and not of $\diz_A$ as in the previous
sections.
In particular for gauge fields with a nonvanishing instanton number
the zero-mode subtraction in \refs{conv1} is always necessary and one deals with
zero-mode subtracted heat kernels
\begin{equation}
\begin{split}
K'(t,A;x)&=K(t,A;x)-\varrho_0(x)=\sum_{p>0} e^{-\mu_p t}\varrho_p(x),\\
\hbox{with}\quad&K(t,A;x)=\langle x\vert
\tr\exp\left(-t\cD^2_A\right)\vert x\big\rangle,
\end{split}
\label{conv3}
\end{equation}
where the sum extends over all $p$ with $\mu_p>0$.
On the torus the smallest nonvanishing 
eigenvalue $\mu_1$ is strictly positive and the zero-mode subtracted
kernel falls off exponentially,
\eqnl{
K'(t,A;x)\longrightarrow e^{-t\mu_1}\varrho_{1}(x)\mtxt{for} t\to \infty
.}{conv5}
On the other hand, the heat kernel of 
the second order  elliptic operator $\cD_A^2$ has 
the \emph{asymptotic} small-$t$ expansion \cite{Minak}
\eqnl{
K(t,A;x)=\frac{1}{t^{d/2}}\Big\{\sum_{n=0}^N
a_n(x)t^n+\mathcal{O}(t^{N+1})\Big\}, 
}{conv7}
where the Seeley-deWitt coefficients  $a_n(x)$ are gauge invariant
local functions built from the field strength and its covariant derivatives
\cite{Gilkey}. 
Inserting this asymptotic expansion into the Mellin transform
\begin{equation}
\begin{split}
\zeta(s,A;x)&=\frac{1}{\Gamma(s)}\int_0^\infty dt\, t^{s-1}
K'(t,A;x)\\
&=\sum_{p>0} \mu_p^{-s}\varrho_p(x)\quad
(s>d/2)\label{sum1}
\end{split}
\end{equation}
one verifies that the analytic continuation
$\zeta(s,A;x)$ of the sum in \refs{sum1}
is a meromorphic function of $s$ \cite{Schwarz:1979ta}. 
In even dimensions it
has a finite number of simple poles as $s=d/2,\dots,2,1$
with residues $a_0(x),a_1(x),\dots, a_{d/2-1}(x)$.
Further its value at $s=0$ is $a_{d/2}(x)-\varrho_0(x)$
\cite{Blau:1988jp}.
Since the field strength and its covariant derivatives 
transform according to the adjoint transformation we 
conclude that the gauge invariant coefficients $a_n$ 
are invariant under center transformations, such that they
cancel in the center average
\begin{align}
\Sigma^{\prime(-2s)}(x)&=
\frac{1}{\Gamma(s)}\!
\int\! dt\,t^{s-1}\cG'(t,A;x) \nonumber
=\sum_k z^\ast_k\, \zeta(s,\Azk,x)\\
&=\sum_{p>0}\Big(\sum_k z_k^\ast\frac{1}{(^{z_k}\hskip-.4mm\mu_p)^s}
\rhozk_p(x)\Big).\label{sum5}
\end{align}
It follows that $\Sigma'^{(-2s)}(x)$ has actually no
poles in the complex $s$-plane and that
$\Sigma'^{(0)}(x)=-\bar\varrho_0(x)$.
Without summing over the center elements the poles
would not disappear and the sum over $p$ would only
exist for $s>d/2$. But if one first averages over the
center and then sums over the quantum number $p$, then
the last sum in \refs{sum5} exist \emph{for all} $s$.
It is important that one first sums over the center elements
and then over the $p$. For example, one finds
\eqnl{
\int d^d x \,\Sigma'^{(0)}(x)=\sum_{p>0}\sum_k z_k^\ast \int d^dx\,
\rhozk_p(x)=0,
}{sum7}
since all densities $\rhoz_p$ integrate to one.
The same result follows from $\Sigma'^{(0)}(x)=-\bar\varrho_0(x)$
since $\bar\varrho_0$ integrates to zero.

Let us now compare the spectral sums built from eigenvalues and densities of
$\cD_A^2$ with spectral sums built from those of $\cD_A$.  
An eigenvalue $\mu_p$ of $\cD_A^2$ is the square of 
an eigenvalue $\lam_p$ of  $\cD_A$. In the \emph{massless} case $\pm\lam_p$ 
are both eigenvalues of the Dirac operator and using $\{\cD_A,\gamma_5\}=0$ it
follows that the eigenvalue densities of $\cD_A$ to positive and
negative eigenvalues are the same,
$\hat\varrho_+=\hat\varrho_-=\hat\varrho$ where $\hat\varrho$ is the
density \refs{sun12}
with eigenfunctions of $\cD_A$. Therefore we have
$\varrho=2\hat\varrho$ for the eigenvalue density of $\cD_A^2$. 
Thus we can conclude for the spectral sums of $\cD_A$
\begin{align}
	\hat\Sigma^{\prime(-2s)}(x)&=\sum_{p>0}\sum_k \nonumber
	z_k^\ast\big((\lamzk_p)^{-2s}
	+(-\lamzk_p)^{-2s}\big){^{z_k}\hskip-.4mm\hat{\varrho}_p}\\
	&=\frac12\sum_{p>0}\sum_k z_k^\ast\big(1+(-1)^{2s}\big)\nonumber
	\abs{^{z_k}\hskip-.4mm\mu_p}^{-s}{\rhozk_p(x)}\\
	&=\frac{1}{2}\big(1+(-1)^{2s}\big)\Sigma^{\prime(-2s)}(x).
	\label{conv18}
\end{align}  
From this it is clear that in the massless case the spectral sum of
$\cD_A^{-2s}$
exists in case the spectral
sum of $(\cD_A^2)^{-s}$ exists and the spectral sums for $\cD_A^{-s}$ vanish
for odd $s$. We conclude that also
the spectral sums
\eqnl{
\hat{\Sigma}^{\prime (-s)}(x)=
\sum_{p>0}\Big(\sum_k z_k^\ast\frac{1}{(^{z_k}\hskip-.4mm\lambda_p)^s}
\,{^{z_k}\hskip-.4mm\hat{\varrho}_p}(x)\Big),
}{conv19}
where again the zero-mode contribution is omitted,
exist for all $s$ and gauge potentials $A_\mu$.
\section{The Schwinger Model at finite Temperature}
For the Abelian instantons on the torus introduced by
\mbox{'t Hooft} \cite{'tHooft:1981ht} all eigenmodes
of the operator $\cD_A^{\;2}$ can
be constructed in the massless limit. The calculations
for Abelian and non-Abelian gauge theories are very similar and  so are the
calculations in
two and four dimensions.
In this section we shall compute the spectral sums
for all instanton configurations of the Schwinger model.
We shall prove that the identity \refs{sl7} holds true with a 
finite constant.
\subsection{Instantons and excited modes on the torus}
The $U(1)$-gauge fields on the two-dimensional torus
fall into topological sectors characterized by
the instanton number $q$. We choose a trivialization
of the $U(1)$-bundles such that in a given sector the
fermionic field satisfies the 'boundary conditions'
\begin{equation}
\begin{split}
\psi(x_0+\beta,x_1)&=-\psi(x_0,x_1),\\\psi(x_0,x_1+L)
&=e^{i\gamma(x)}\psi(x_0,x_1),\, \gamma=-\frac{2\pi q}{\beta}x_0,
\end{split}
\label{sch1}
\end{equation} with $q\in\Z$.
The gauge potentials are periodic up to a gauge transformation,
\eqnl{
A_\mu(x_0,x_1+L)-A_\mu(x_0,x_1)=\partial_\mu\gamma(x).}{sch5}
The fields with minimal Euclidean action have constant
field strength. We shall calculate spectral sums for
the instanton solutions
\eqnl{
A_0=-\frac{\Phi}{V} x_1+\frac{2\pi h}{\beta},\quad
A_1=0\mtxt{with}F_{01}\equiv B=\frac{\Phi}{V},}{sch7}
where $V=\beta L$ is the volume of spacetime and $h$
an arbitrary constant. The flux $\Phi$ of $B$ 
is related to the instanton number by $\Phi=2\pi q$.
Without loss of generality we assume that the integer 
$q$ is positive. The eigenvalues and 
$q$ ground states of the massless Dirac operator
$\cD_A=i\gamma^\mu(\partial_\mu-iA_\mu)$
have been calculated earlier in \cite{Sachs:1991en}.
Here we shall construct all excited modes of $\cD_A$.

The twisted gauge potential \refs{un3} is equal
to the untwisted potential with shifted $h$,
\eqnl{
\alphaA_0=-B x_1+\frac{2\pi}{\beta}(h+\alpha),
\quad \alphaA_1=0.}{sch11}
Hence it will do to study the spectral problem for
$A_\mu$ in \refs{sch7}. The spectral sums will be
compared with the Polyakov loop variable
\eqnl{
P(x_1)=e^{2\pi ih-i\Phi\, x_1/L}.
}{sch13}
For the instanton potential the straight and all dressed Polyakov 
loops with base $(0,x_1)$ have the same 
value \refs{sch13}. Then \refs{sl5} implies that the
spectral sums must have the form \refs{sl7}.

Below we shall calculate spectral sums for the 
squared Dirac operator $\cD_A^{\,2}$ with vanishing
mass. For a positive instanton number its eigenvalues
$\mu_p=\lam_p^2$
are \cite{Sachs:1991en}
\eqnl{
\mu_p=\left\{\begin{array}{ll}
                 0&\text{degeneracy: } q \\
                 2p B
&\text{degeneracy: }2 q.
                 \end{array}\right.
}{sch15}
Since they are independent of $h$ they are also invariant
under center transformations parameterized by $\alpha$.
It follows that $\Tr f(\alphacD)$ is independent of $\alpha$ such
that the spectral sums $\cS_f(A)$ in \refs{un7} vanish for all functions $f$.
Since the spatial average of $P(x_1)$ vanishes as
well, this corroborates the conjectured result \refs{sl7}.

For the instantons \refs{sch7} the spectral problem reads
\eqnl{
\cD_A^{\;2}\psi_p=-\left(D_\mu D^\mu+\gamma^0\gamma^1 F_{01}
\right)\psi_p=\mu_p\psi_p
}{sp1}
with boundary conditions \refs{sch1}. We choose a chiral
representation with $\gamma^0\gamma^1=\sigma_3$
and diagonalize $i\pa_0$. On the cylinder
$[0,\beta]\times \R$ the
antiperiodic eigenmodes read
\eqnl{
\chi_{p,\ell}(x)=e^{-i\pi x_0/\beta}e^{2\pi i\ell
x_0/\beta}\,\xi_p(x_1),\quad(\ell\in\Z)
}{sp3}
with time-independent mode functions $\xi_p$.
These functions must solve the Schr\"odinger equation for the supersymmetric 
harmonic oscillator
\eqnl{
\left(-\pa_y^2+B^2 y^2-B\sigma_3\right)\xi_p
=2pB\xi_p,
}{sp5}
where $y$ is the shifted spatial coordinate
\eqnl{
y=x_1+\frac{L}{q}\left(\ell-h-1/2\right).}{sp7}
Eigenfunctions with $\sigma_3\xi_p=\xi_p$ are called
right-handed. Every right-handed eigenmode with energy 
$2p B$ gives rise to a left-handed eigenmode $\sigma_2\xi_p$
with energy $(2p+1)B$. Hence we may as well focus on the 
right-handed sector. The zero energy states are
\begin{equation}
\begin{split}
\chi_{0,\ell}(x)&=e^{-i\pi x_0/\beta}e^{2\pi i \ell x_0/\beta}
\xi_0(y),\\ \xi_0(y)&=\left(\frac{B}{\pi}\right)^{1/4}\!
e^{-B y^2/2},
\end{split}\label{sp9}
\end{equation}
where $y(x_1)$ has been defined in \refs{sp7}. The
excited eigenmodes contain Hermite polynomials
\eqnl{
\chi_{p,\ell}(x)=c_p H_p(\sqrt{B}\, y)\,\chi_{0,\ell}(x),\quad
c_p^2=\frac{1}{2^p p!}.}{sp11}
Here we consider only right-handed modes and identify the
nonvanishing component of a right-handed solution with
the solution itself.
The modes \refs{sp11} do not satisfy the boundary conditions \refs{sch1} since
\eqnl{
\chi_{p,\ell}(x_0,x_1+L)=e^{i\gamma(x)}\chi_{p,\ell+q}(x_0,x_1),}{sp13}
but the true orthonormal eigenfunctions on the
\emph{torus} are just superpositions of these functions
\eqnl{
\psi_{p,\ell}(x)=\sum_{s} \chi_{p,\ell+sq}(x),\quad \ell=1,\dots,q.}{sp15}
Note that the eigenvalues do not depend on $\ell$.
Recalling the $\ell$-dependence of $y$ the modes read
\begin{multline}
\psi_{p,\ell}(x)=\frac{c_p}{\sqrt{\beta}}\,
 e^{2\pi i (\ell-1/2) x_0/\beta}\\\times
\sum_{s}H_p\left(\sqrt{B}\, (y+sL)\right)\,\xi_0(y+sL)\,
e^{2\pi i sq x_0/\beta}.
\label{sp17}
\end{multline}
We give a second representation which can be obtained
by a Poisson resummation. In the appendix
we show that the eigenmodes have the alternative
representation
\begin{multline}
\psi_{p,\ell}(x)
=\frac{i^pc_p}{\sqrt{q}\sqrt{L}}\, e^{2\pi i(\ell-1/2-qy/L)x_0/\beta}\\\times
\sum_s H_p\left(\sqrt{B}(x_0+s\beta/q)\right) 
\xi_0\left(x_0+s\beta/q\right)e^{-2\pi is y/L}.
\label{sp19}
\end{multline}
For $p=0$ both sums define $\theta$-functions and one
recovers their modular transformation property.
\subsection{Spectral sums}
To calculate arbitrary spectral sum densities 
$\langle x\vert \tr f(\cD_A^2)\vert x\rangle$
we determine the density of the eigenvalue $\mu_p$ in
the \emph{right-handed sector},
\eqnl{
\varrho^+_p(x)=\sum_{\ell=1}^q \vert\psi_{p,\ell}\vert^2.
}{den1}
The density in the left-handed sector is
$\varrho^-_p=\varrho^+_{p-1}$. Henceforth we shall
skip the superscript $+$. To compute the sums over $\ell$
we use the representations \refs{sp19} for the eigenmodes,
since in this form they show a simple dependence 
on  the quantum number $\ell$ (recall 
that $y\propto \ell$). Using
\begin{multline}
\sum_{\ell=1}^q e^ {2\pi i (s-r) y/L}\\=q\delta_{s-r,nq}\,
e^{2\pi i(s-r)\{x_1/L-(h+1/2)/q\}},\, n\in \Z,\label{den3}
\end{multline}
the sum over $\ell$ can be carried out.
Twisting with $\alpha$ as in \refs{sch11} amounts to shifting
$h$ by $\alpha$. Since the eigenvalues do not see 
the twist we may first calculate the integral of $\,^{\alpha}\!\varrho_p$
over the center,
\eqnl{
\bar\varrho_p(x)=\int_0^1 d\alpha\,e^{-2\pi i\alpha} \;\,{^\alpha}\hskip-.6mm\varrho_p(x)}{den5}
and afterwards sum over $p$ to calculate the spectral sums \refs{un7}.
The $\alpha$-dependence of $\,^{\alpha}\!\varrho_p$ in \refs{den1}
comes only from the exponential factor in \refs{den3} and the 
corresponding integral over the center elements,
\begin{multline}
\int d\alpha\, e^{-2\pi i\alpha}e^{2\pi i(s-r)\{x_1/L-(h+\alpha+1/2)/q\}}\\
=-P(x_1)\,\delta_{s-r,-q}\label{den7}
\end{multline}
is proportional to $P(x_1)$.
Due to the constraints imposed by the Kronecker symbol
the double sum for $\bar\varrho_p$, resulting from the series 
representation \refs{sp19} for the eigenmodes, reduces to one sum
\eqnl{
\bar\varrho_p(x)=c\cdot \sigma_p (x_0)\cdot P(x_1)\mtxt{with}
c=-\frac{1}{L}\left(\frac{B}{\pi}\right)^{1/2},
}{den9}
where the series $\sigma_p$ only depends on $x_0$.
Each term in the series contains a product of two
Hermite polynomials $H_p$ and two Gaussian functions
$\xi_0$. Defining the variable  $x_s=\sqrt{B}\{x_0+\beta(1/2+s/q)\}$
the series takes the form
\begin{multline}
\sigma_p(x_0) =c_p^2\,e^{-B\beta^2/4}
\sum_s H_p\left(x_s-\sqrt{B}\beta/2\right)\\\times
H_p\left(x_s+\sqrt{B}\beta/2\right) e^{-x_s^2}.
\label{den11}\end{multline}
To integrate over time we observe that $\sigma_p$
is periodic in $x_0$ with period $\beta/q$ such that
\eqnn{
\int_0^\beta dx_0 \sum_s \dots =q\int_0^{\beta/q}\sum_s\longrightarrow
q\int_{-\infty}^\infty.}
Now we can apply the integral formula \cite[(7.377)]{Gradshteyn}
\eqnl{
\int_{-\infty}^\infty dx\, H_p(x+y)H_p(x+z)\,e^{-x^2}=\frac{\sqrt{\pi}}{c_p^2}
L_p(-2zy),}{den12}
where $L_p$ denotes the Laguerre polynomial of order $p$,
and this leads to
\eqnl{
\int dx_0\,\sigma_p(x_0)=
q\sqrt{\frac{\pi}{B}}\, L_p(B \beta^2/2)\,
e^{-B \beta^2/4}.}{den13}
Inserting this result into \refs{den9} yields
\begin{equation}
\begin{split}
\bar\varrho_p(x_1)&\equiv\int dx_0\,\bar\varrho_p(x)\\&=
-\frac{q}{L}P(x_1)\, L_p(\pi q\tau)\, e^{-\pi q\tau/2},\quad
\tau=\frac{\beta}{L}.
\end{split}
\label{den15}
\end{equation}
Taking the trace in spinor space amounts to adding the
contributions of the right- and left-handed
sectors. This finally leads to the following explicit result for the 
spectral sums in \refs{un7}
\begin{multline}
\cS_f(A;x_1)=-\frac{q}{L}\,P(x_1)\\\times
\sum_{p=0}^\infty f(\mu_p)\,\big\{L_p(\pi q\tau) 
+L_{p-1}(\pi q\tau)\big\}\,e^{-\pi q\tau/2},
\label{den17}
\end{multline}
where we defined $L_{-1}=0$. This is the main result of this
section. As expected on general grounds every function 
giving rise to a convergent
series \refs{den17} defines a spectral function $\cS_f(A,x_1)$
which is proportional to the Polyakov loop. How fast
the series converges to the asymptotic value depends on
the particular choice of $f$.

\paragraph*{Gaussian sum:}
Here we consider the Gaussian spectral sum
\eqnl{
\cG(A;x_1)=\int dx_0\,\int d\alpha\,  e^{-2\pi i\alpha}\langle x\vert
\,\tr\exp\left(-\alphacD_A^2\,/\mu^2\right)\vert x\rangle
}{gs1}
with some mass parameter $\mu$. The integrand $\langle x|\ldots|x\rangle$ is just
the heat kernel of $\alphacD^2$ on the diagonal in position space.
The resulting series \refs{den17} with $f(\mu_p)=\exp(-2Bp/\mu^2)$
relates to the generating function for the Laguerre 
polynomial \cite[10.12 (17)]{Erdelyi}
\eqnl{
\frac{1}{1-z}\exp\left(-\frac{xz}{1-z}\right)=
\sum_{p=0}^\infty L_p(x)z^p,\quad \abs{z}<1.}{gs3}
One obtains the simple result
\eqnl{
\cG(A;x_1)=-\frac{q}{L}\coth\frac{B}{\mu^2}\,
\exp\left(-\frac{\pi q\tau}{2} \coth\frac{B}{\mu^2}\right) P(x_1).
}{gs5}
Using $B=2\pi q/\beta L$ one finds for $L\gg qT/\mu^2$ that the
relation between the Gaussian spectral sums and
the Polyakov loop is the same in all instanton sectors,
\eqnl{
\cG(A;x_1)\stackrel{L\to\infty}{\longrightarrow}
-\frac{\mu}{2\pi}(\mu\beta)\,e^{-(\mu\beta)^2/4}\,P(x_1).
}{gs7}
A natural energy scale at finite temperature would
be the temperature itself, $\mu=T$. With this choice
the infinite-volume result simplifies further,
\eqnl{
{\cG}_{\infty}(A;x_1)=-\frac{T}{2\pi e^{1/4}}\,P(x_1)\mtxt{for}
\mu=T,\;\; L\gg \beta.}{gs9}
\paragraph*{Propagator sum:}
Here we consider the propagator sum
\eqnl{
\Sigma^{(-2)}(A;x_1)=\int dx_0\,\int d\alpha\, e^{-2\pi i\alpha}\left\langle x\big\vert
\,\tr'\left(\alphacD_A\right)^{-2}\big\vert x\right\rangle,
}{prop1}
where $\tr'$ means the trace without singular contribution 
of the $q$ zero-modes.
Making use of the summation formulas \cite[(8.976)]{Gradshteyn}, \cite{Hall} 
\eqnl{
\sum_{p=1}^\infty \frac{L_p(x)}{p}=-\gamma-\log x\mtxt{and}
\sum_{p=1}^\infty \frac{L_{p-1}(x)}{p}=e^x\Gamma(0,x)}{prop3}
the spectral sum is given in terms of the Euler constant
$\gamma$ and the incomplete Gamma-function,
\eqnl{
\Sigma^{(-2)}=\frac{\beta}{4\pi}
\big\{\gamma+\log(\pi q\tau)-e^{\pi q\tau}\,\Gamma(0,\pi q\tau)\big\}\,
e^{-\pi q\tau/2}\,P(x_1).}{prop5}
In the large volume limit $\tau=\beta/L$ tends to zero and one
obtains the simpler relation
\eqnl{
\Sigma^{(-2)}\stackrel{\beta/L\to\infty}{\longrightarrow}
\frac{\beta}{2\pi}(\gamma+\log (\pi q\tau))\cdot P(x_1)+O(\beta/L).
}{prop7}
In two dimensions $\Tr\,\cD_A^{-2}$ is logarithmically divergent 
in the ultraviolet for \emph{all} background fields. For the instanton 
potential with $\mu_p\propto p$ this is evident. On the other hand,
the spectral sum $\Sigma^{(-2)}$ is finite. Integrating over the 
center removes the divergence. 

\begin{figure}[ht]
\includegraphics[width=8.6cm]{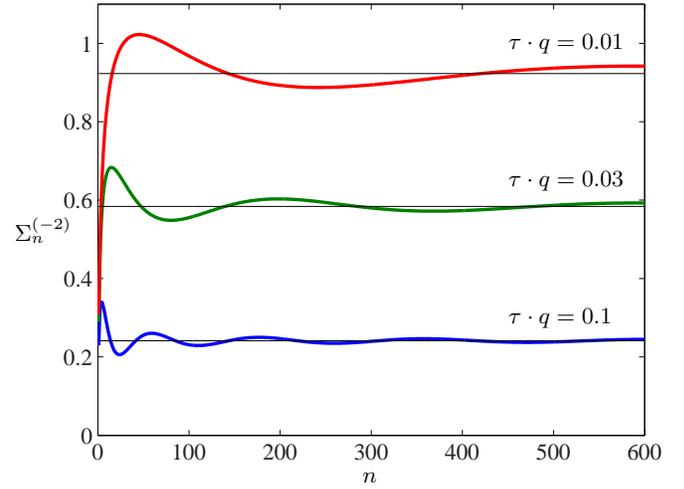}
\caption{\label{fig:PropagatorU1} 
Partial  propagator sums $\Sigma^{(-2)}_n$ for $P(x_1)=1$ and their limiting values
for
different torus parameters ($\beta=1$).}
\end{figure}

Figure~\ref{fig:PropagatorU1} shows  the partial propagator
sums 
\eqnl{\Sigma^{(-2)}_n=\frac{q}{L}\sum_{p>0}^n\frac{1}{\mu_p}\,\big\{
L_p(\pi q\tau) 
+L_{p-1}(\pi q\tau)\big\}\,e^{-\pi q\tau/2}}{prop9} for $P(x_1)=1$ as a function
of the
included
eigenmodes and their limiting values calculated from~\refs{prop5}. Depending on
the ratio $\tau$ and the
instanton number $q$ they converge fast to their limiting values. 

\section{Finite Temperature $\mathbf{SU(2)}$ Gauge Theory}
In this section we calculate all eigenvalues and eigenfunctions
of the Dirac operator for twisted and untwisted instanton configurations 
with  constant field strength on the torus $\T^4=[0,\beta]\times [0,L]^3$ with 
volume $V=\beta\cdot V_s$. As a result we obtain explicit
expressions for the spectral densities. 
As expected, the Gaussian sums reproduce $P(\mbx)$
and get their main contribution from small eigenvalues.
\subsection{Instantons with constant field strength}
Following t'Hooft \cite{'tHooft:1981ht} we consider configurations
with constant field strength,
\eqnl{
A_0=\left(-Ex_3+\frac{2\pi h_0}{\beta}\right)\sigma_3,\;
A_2=Bx_1\sigma_3
,\;
A_1=A_3=0}{na1}
and assume that the constant chromo-electric and chromo-magnetic field 
components $E$ and $B$ are positive. The instanton number is proportional to
$EB$ and to the volume of space-time,
\eqnl{
q=\frac{1}{32\pi^2}\int_{T^4}\varepsilon_{\mu\nu\alpha\beta}\,\tr F_{\mu\nu}F_{\alpha\beta}
=\frac{EB}{2\pi^2}V,}{na3}
and the Polyakov loop is periodic in the electric field,
\eqnl{
P(\mbx)=2\cos\left(2\pi h_0-E\beta x_3\right).}{na4}
For $q>0$ the gauge potential is 
periodic up to a nontrivial gauge transformation, 
$A_\mu(x+L_\nu e_\nu)=A_\mu(x)+\partial_\mu\gamma_\nu(x)$,
with transition functions given by 
\eqnl{
\gamma_1(x)=BL x_2 \sigma_3\mtxt{and}
\gamma_3(x)=-ELx_0\sigma_3.}{na5}
The fermions are antiperiodic in time, periodic in $x_2$ and  fulfill
\begin{equation}
\begin{split}
\psi(x+Le_1)&=e^{i\gamma_1(x)}\psi(x),\\
\psi(x+Le_3)&=e^{i\gamma_3(x)}\psi(x).
\end{split}
\label{na7}
\end{equation}
Consistency demands that the fluxes in the $03$ and $12$ planes are
both quantized,
\eqnl{
\Phi_{03}=E(\beta L)=2\pi q_{03},\quad
\Phi_{12}=BL^2=2\pi q_{12}}{na9}
 with $q_{03},q_{12}\in\Z$
such that the instanton number $q=2q_{03}q_{12}$ is always even. 
In the chiral representation with
\eqnl{
\gamma^0=\begin{pmatrix}0&i\sigma_0 \cr-i\sigma_0&0
         \end{pmatrix}
,
\gamma^i=\begin{pmatrix}0&\sigma_i\cr\sigma_i&0\end{pmatrix},
\gamma^5=\begin{pmatrix}\sigma_0&0\cr 0&-\sigma_0\end{pmatrix},}{na11}
where $\sigma_0$ is the two-dimensional unit matrix, the squared Dirac operator
takes the form
	\eqnl{
\cD_A^2=-D^2-\sigma_3\otimes \begin{pmatrix}(B+E)\sigma_0&0\cr
0&(B-E)\sigma_0\end{pmatrix}}{na13}
with $D^2=D_\mu D^\mu.$
The Pauli term acts with $\sigma_3$ on the color-$SU(2)$-indices,
with $(B+E)\sigma_0$ on right-handed spinors
and with $(B-E)\sigma_0$ on left-handed spinors.
$D^2$ is proportional to the four-dimensional 
identity in spinor space and commutes with 
$\pa_0$ and $\pa_2$. The (anti)periodic
eigenfunctions decaying in the $x^1$ and $x^3$ directions have the form
\eqnl{
\psi_{p,\ell}(x)=e^{-\pi ix_0/\beta}
e^{2\pi i (\ell_0x_0/\beta +\ell_2x_2/L)}\xi_p(x_1,x_3).
}{na15}
On the functions $\xi_p$ the operator $-D^2$ reduces to the sum 
of two commuting harmonic oscillator Schr\"odinger operators,
one acting on $x_3$ and the other on $x_1$.
On an  eigenfunction of $\sigma_3$ in color space
the operators read
\begin{equation}
\begin{split}
H_{03}&=(-\pa_y^2+E^2y^2)\otimes\id,\\
H_{12}&=(-\pa_z^2+B^2z^2)\otimes\id,
\end{split}\label{na17}
\end{equation}
where we introduced the shifted coordinates $(y,z)$ as follows,
\eqnl{
y=x_3-\frac{L}{q_{03}}\left\{h_0\mp \ell_0\pm \ft12\right\},\quad
z=x_1\mp\frac{L}{q_{12}}\ell_2}{na23}
for
$\sigma_3\xi_p=\pm\xi_p.$
Thus we recover two copies of the Schwinger model and conclude
\eqnl{
\psi_{p,\ell}(x)=e^{-\pi ix_0/\beta}
e^{2\pi i (\ell_0x_0/\beta +\ell_2x_2/L)}\xi_{p_3}(y)\xi_{p_1}(z)\chi,
}{na19}
where $\chi$ is a constant right- or left-handed spinor and
eigenvector of $\sigma_3$ in color space.
The $\xi$'s are eigenfunctions of harmonic oscillator operators,
\begin{equation}
\begin{split}
H_{03}\xi_{p_3}&=2(p_3+1)E\xi_{p_3},\\
H_{12}\xi_{p_1}&=2(p_1+1)B\xi_{p_1}.\label{na21}
\end{split}
\end{equation}
The eigenfunctions on the torus $\T^4$ fulfilling the 'boundary
conditions' \refs{na7} are superpositions of the eigenmodes
\refs{na19} and read
\eqnl{\psi_{p,\ell}(x)=\psi_{p_3,\ell_0}(x_0,x_3)\,
\psi_{p_1,\ell_2}(x_2,x_1)\,\chi,}{na25}
with $\ell_0\in\{1,\dots,q_{03}\}$  and $\ell_2\in\{1,\dots,q_{12}\}$.
The explicit form of the factors are given in \refs{sp17} or
\refs{sp19} with obvious replacements.
For every pair of quantum numbers $p=(p_3,p_1)$
there are $q_{03}\cdot q_{12}$ eigenmodes of the squared
Dirac operator.

For $\sigma_3\chi=\chi$ in color space the variables $y$ and $z$ are given in
\refs{na23} with the upper signs and in spinor space the squared Dirac operator 
acts on the modes in \refs{na25} as follows:
\eqnl{
\cD_A^2\to 2\left(p_3E+p_1B\right)\id_4+
2\begin{pmatrix}0&0\cr 0&E\sigma_0\end{pmatrix}
}{na27}
For $\sigma_3\chi=-\chi$ the variables $y$ and $z$ are given in
\refs{na23} with the lower signs and
\eqnl{
\cD_A^2\to 2\left(p_3E+p_1 B+B\right)\id_4+
2\begin{pmatrix}E\sigma_0&0\cr 0&0\end{pmatrix}.
}{na29}
For every pair $(p_1,p_3)$ and generic $E,B$ there are four 
eigenvalues, each with degeneracy $q=2q_{03}\cdot q_{12}$.
In particular, there exist $q$ right-handed zero-modes 
in agreement with the index theorem.
\subsection{Spectral sums}
To calculate the densities 
$\varrho_p(x)=\sum_{\ell}\vert\psi_{p,\ell}(x)\vert^2$
we use the representation \refs{sp19} for the factors $\psi_{p_3,\ell_0}$
and $\psi_{p_1,\ell_2}$ in \refs{na25}. Then
the sums over $\ell_0$ and $\ell_2$ are calculated similarly
as for the Schwinger model. The densities $\langle x\vert \tr f(\cD_A)\vert x\rangle$
do not depend on $x_2$ since $\cD_A$ commutes with
$\pa_2$. Hence we may as well average over
the $x_2$-coordinate.

This leads to a contribution 
\eqnl{
\varrho_{p_3}(x_3)\varrho_{p_1}(x_1)=
\int dx_0\varrho_{p_3}(x_0,x_3)\cdot
\frac{1}{L}\int dx_2\varrho_{p_1}(x_1,x_2)}{nas3}
of the $q_{03}\cdot q_{12}$ eigenmodes  with fixed $(p_3,p_1)$ and
fixed $\chi$. The explicit form of the factors is
\begin{align}
&\varrho_{p_3}(x_3)=\frac{q_{03}}{L}\nonumber
\\&\times\Big(1+\sum_{n=1}^\infty 
 (-)^n \tr \cP^n(\mbx)\, L_{p_3}\left(\frac{E\beta^2n^2}{2} \right) e^{-E(\beta
n/2)^2}\Big),
\label{nas5a}\\
\nonumber
&\varrho_{p_1}(x_1)=\frac{q_{12}}{L^2}\\&\times\Big(1+
\sum_{n=1}^\infty \tr \cQ^n(\mbx)
L_{p_1}\left( \frac{BL^2n^2}{2} \right) e^{-B(L n/2)^2}\Big),\label{nas5b}
\end{align}
where we introduced the Wilson-loop winding once around
the $x^2$-direction,
\eqnl{
\cQ(\mbx)=e^{i\int A_2 dx_2}=\hbox{diag}\big( e^{iBx_1L},e^{-iBx_1L}\big).
}{nas6}
Note that the factors do not depend on the
choice of signs in \refs{na23} and this simplifies the analysis
considerably. So far we did not sum or integrate over the twists and 
this is the reason why loops winding an arbitrary number of times
around the time direction contribute to the
sum \refs{nas5a}.

The eigenvalues of $\cD_A^2$ are given by the eigenvalues of the
matrices in (\ref{na27}, \ref{na29}). It is not difficult to see that 
eigenfunctions with the following values of $(p_3,p_1)$
contribute to a given eigenvalue $\mu_{a,b}=2(a E+bB)$
of $\cD_A^2$ with $a,b\geq 0$:
\eqnl{
\mu_{a,b}\Rightarrow
(p_3,p_1)=(a,b),(a-1,b),(a,b-1),(a-1,b-1).}{nas7}
Generically there exist $4q$ eigenfunctions with the same eigenvalue $\mu_{a,b}$.
But for $a=0$ or $b=0$ there exist only $2q$ eigenfunctions
and for $a=0$ and $b=0$ there exist $q$ zero-modes.
For a given eigenvalue $\mu_{ab}$ we have the following
densities ($a,b\geq 1$)
\begin{equation}
\varrho_{a,b}(\mbx)=2\big\{\varrho_{a}(x_3)
+\varrho_{a-1}(x_3)\big\}
\big\{\varrho_b(x_1)+\varrho_{b-1}(x_1)\big\}\label{nas13}
\end{equation}
where we defined $\varrho_{-1}=0$.
\paragraph*{Twisting in $SU(2)$:}
Twisting the gauge potential inside the gauge group $SU(2)$ amounts to
adding $1/2$ to $h_0$ or equivalently changing the sign
of the Polyakov loop. The density $\varrho_{b}(x_1)$
is unchanged but
\begin{multline}
\bar\varrho_{a}(x_3)
-{^{z}\!\varrho_{a}}(x_3)=
-\frac{q_{03}}{L}
\\\times\sum_{n=1,3,5,\dots}  \tr
\cP^n(\mbx)\,
L_{a}\big(\frac{ n^2\beta^2E}{2}\big) \,e^{-E(n\beta/2)^2}.\label{nas15}
\end{multline}
Only odd powers of the untraced Polyakov loop contribute
to the sum $\varrho+z^\ast\,^{z}\hskip-.5mm\varrho\,$
over the center elements of $SU(2)$. This confirms with our
general analysis given earlier.
\paragraph*{Twisting in $U(2)$:}
We may twist the $su(2)$ gauge potential with center elements
of $U(2)$ or equivalently twist the boundary conditions
by an arbitrary phase factor $e^{2\pi i\alpha}$. Averaging over 
the phases as in \refs{den5} and below leads to
\eqnl{
\int d\alpha \;{^\alpha}\hskip-.6mm\varrho_{a}(x_3)\,e^{-2\pi i\alpha}
=-\frac{q_{03}}{L}\, P(\mbx)\,
L_{a}\big( \beta^2E/2\big) \,e^{-E(\beta/2)^2}.}{nas19}
Integrating over the center of $U(2)$ the sum over all windings $n$
collapses to the contribution with just one winding which is
proportional to $P(\mbx)$. Again this corroborates with our general
analysis.
\paragraph*{Gaussian sum:}
To calculate the Gaussian spectral sum for the instanton configurations,
\eqnl{
\cG(A;\mbx)=\int dx_0\,\sum_k z^\ast_k\, \langle x\vert
\,\tr\exp\left(-{^{z_k}}\hskip-.5mm \cD_A^2\,/\mu^2\right)\vert x\rangle
}{gsna1}
one needs to calculate the sums $\sum_{a,b}\bar\varrho_{a,b}\exp(-\mu_{a,b}/\mu^2)$
with density \refs{nas13}. Summing over the $SU(2)$-center or integrating
over the $U(2)$-center amounts to replacing $\varrho_a$ 
in \refs{nas13} by the difference \refs{nas15}
or the integral \refs{nas19}. Again the resulting series are calculated with 
the help of \refs{gs3}. After summing over the $SU(2)$-center elements
one obtains
\begin{multline}
\cG(A;\mbx)=
\frac{q}{2V_s}\coth\frac{E}{\mu^2}\coth\frac{B}{\mu^2}\\\times\negthickspace
\sum_{n=1,3,5,\dots}\negthickspace(-)^n\tr \cP^n(\mbx)\, \exp\left(-\frac{\pi q_{03}\tau n^2}{2}
\coth \frac{E}{\mu^2}\right)\\
\times \left[1+\negthickspace\sum_{n'=1,2,\dots}\negthickspace\tr \cQ^{n'}(\mbx) \exp\left(-\frac{\pi
q_{12}n'^2}{2}
\coth \frac{B}{\mu^2}\right)\right].\label{gsna3}
\end{multline}
Not unexpected the first sum contains the center- and gauge
invariant variables $\tr \cP,\tr\cP^3,\tr\cP^5\dots$.
If the spatial extend of the torus becomes large
and we fix $q_{03}$ and $q_{12}$, in which
case the fields $E$ and $B$ tend to zero, then
the Gaussian sum simplifies to
\begin{equation}
\begin{split}
{\cG}_{\infty}(A;\mbx)&=\frac{\mu^4\beta}{4\pi^2}
 \sum_{n=1,3,\dots}(-)^n
\tr\, \cP^n(\mbx)\cdot e^{-(\mu\beta n/2)^2}
\label{gsna5}
\end{split}
\end{equation}
for $L\gg \min\{qT/\mu^2,\, q/\mu\}$.
On the other hand, for fixed $q_{03}, q_{12}$ and fixed spatial extend 
 we regain the the zero-mode contributions to~\refs{gsna3} for
$\mu^{-2}\to\infty$. For $\mu^{-2}\to0$ we recover ${\cG}_\infty$ in
\refs{gsna5}. This implies an exponential decay with $\mu^{-2}$ as proposed in
the general discussion on the convergence of the spectral sums. 

If we allow for $U(2)$-twists with arbitrary phase factors
then the resulting Gaussian sum is again given by the
formula \refs{gsna3}, but in the first sum over $n$ only
the term with $n=1$ contributes. In the thermodynamic 
limit $L\to\infty$ with fixed fluxes we find the simpler result
\eqnl{
{\cG}_{\infty}(A;\mbx)=-\frac{\mu^3}{4\pi^2}\, \mu \beta
\, e^{-(\mu\beta)^2/4}\,P(\mbx).}{gsna7}
As expected, twisting with arbitrary center elements in $U(2)$ 
removes the higher powers of the Polyakov loop. The formula 
is almost identical to the result \refs{gs7} for the Schwinger model.

\paragraph*{Propagator sums}
The propagator sums are not absolutely convergent and the summation has to
be carried out over fixed energy shells. Thus  they are defined as
\eqnl{\Sigma^{\prime(-2s)}(x)=\lim_{\Lambda\to\infty}\sum_{\substack{a,b\\0<\mu_
{ ab } \leq\Lambda } } \frac { 1 } { \mu_{ab}^{s}} \bar\varrho_ {a,b}.
}{propna1} The existence of the right hand side for $s>0$ can explicitly be
shown with the Mellin transformation of the zero mode subtracted
heat-kernel~\refs{gsna3} in accordance with the general
discussion following \refs{sum5}.
\section{Numerical investigation  }\label{Numerical}

\subsection{ Numerical setup}

\begin{figure*}[t]
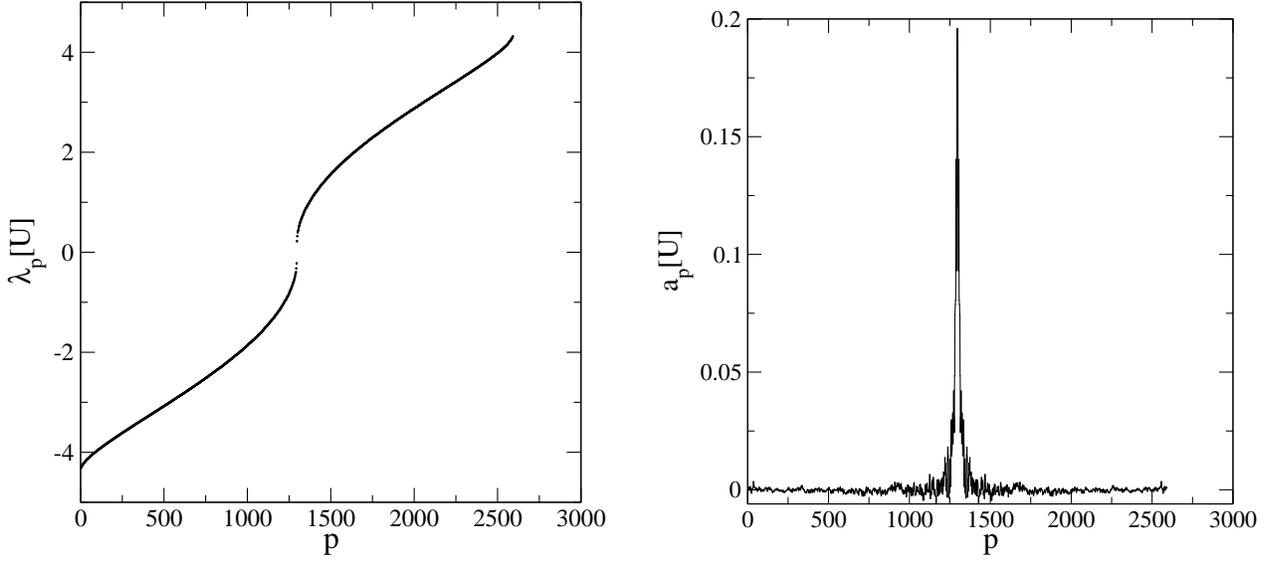

\centerline{
\includegraphics[width=8cm]{eigenv.eps} \hspace{0.5cm}
\includegraphics[width=8cm]{sur_en.eps}
}
\caption{ All eigenvalues of the staggered quark operator for
a particular lattice configuration on a $6^4$ lattice (left).
Contribution to the interface energy as function of the mode index
(right).
 }
\label{fig:k1}
\end{figure*}
Our lattice gauge simulations were carried out on a $N_t\times N_s^3$
lattice using an improved action which is optimized for good
rotational symmetry and good scaling~\cite{Langfeld:2007zw}. We
confined ourselves to the gauge group $SU(2)$ and to a limited range of
lattice spacings in this first exploratory study. The action is given by
\eqnl{
S = \beta  \sum _{\mu > \nu, x} \Bigl[ \gamma _1 
P_{\mu \nu }(x)  +  \gamma _2  P^{(2)}_{\mu \nu }(x)
\Bigr]  , }{eq:k1}
where $ P_{\mu \nu }(x) $ is the standard plaquette expressed
in terms of the link fields $U_\mu (x) \in SU(2)$, i.e.,
\eqnn{
P_{\mu \nu }(x)= \frac{1}{2} \, \tr \Bigl[
U_\mu(x) \, U_\nu (x+\mu) \, U^\dagger _\mu (x+\nu) \,
U^\dagger _\nu (x) \; \Bigr]  ,}
and $ P^{(2)}_{\mu \nu }(x) $ is half the trace of the
$2\times 2$ Wilson loop. We used the
parameter set given in Table~\ref{table:ParamterSet}~\cite{Langfeld:2007zw}.
\begin{table}
\caption{\label{table:ParamterSet} Parameter set used for the simulations}
\begin{ruledtabular}
\begin{tabular}{cccc}
$\beta $ & $\gamma _1 $ & $\gamma _2 $ & $\sigma a^2 $ \\ \hline
$1.35$   & $2.0348$ & $-0.10121$   & $0.1244(7)$
\end{tabular}
\end{ruledtabular}
\end{table}
Thereby, $a$ is the lattice spacing and $\sigma $ is the string tension.
For the study of the eigenmodes of the Dirac operator, we  used
the staggered Dirac operator:
\eqnl{
\langle x\vert \cDU\vert y\rangle = \sum _{\mu=0}^3 \eta _\mu (x)  \Bigl[ U_\mu(x)
\, \delta _{x + \mu, y}  -  U^\dagger _\mu (x-\mu)
\, \delta _{x -\mu, y} \, \Bigr] , }{eq:k2}
where the phase factors are given by
$ \eta _\mu(x)=  (-1)^{x_0+\ldots + x_{\mu -1} }.$

\subsection{ Energy of the $\Z_2$ interface}
The group $SU(2)$ has only the two center elements
$\id$ and $z=-\id$.
The center transformation $U \rightarrow {^z}U$ in \refs{lat5}
does not change the (gluonic) action (\ref{eq:k1}). Defining the energy
of the interface by the action difference, $\cA =
\langle S[ {^z}U] - S[ U]\rangle $, there is no penalty
in the action for inserting such an interface.

This situation changes when \textit{dynamical} quarks are included: the
quark determinant
is not invariant under the mapping $U \rightarrow {^z}U$.
Considering the quark determinant as an integral part of the total
action, the energy of the interface is now given by
\eqnl{
\cA = \left\langle \ln \, \frac{ \det {^z}\cDU  }{
 \det \cDU } \, \right\rangle  . }{eq:k4}
Representing each determinant $ \det \cDU$ by the product
of the eigenvalues $\lambda_p$ of the corresponding Dirac operator,
we can equally well write
\eqnl{
\cA = \sum_p \langle a_p \rangle  ,\quad
a_p = \ln\left(\frac{ ^z\lambda_p}{\lambda_p}\right)  . }{eq:k5}
We note that in quenched approximation (quark effects
on the link configurations are neglected), the
surface energy $\cA$ vanishes since the configurations
${^z}U$ and $U$ contribute with equal probability to the Monte-Carlo
average.
Nevertheless it is instructive to study $a_p$ for a single
lattice configuration generated in quenched approximation.
Figure~\ref{fig:k1}, left panel, shows all $2 N_t N_s^3$ eigenvalues for
a particular lattice configuration, while the right panel shows the
contribution $a_p$ to the interface energy. We observe that
the dominant contribution to the interface energy arises from the
low lying eigenmodes of the Dirac operator (Figure~\ref{fig:k1}, right panel).
Our findings also suggest that
the mode sum in (\ref{eq:k5}) converges. This would imply that the
interface energy is entirely determined by the IR regime
of the quark sector.

\subsection{Polyakov loops}

\begin{figure*}[t]
\centerline{
\includegraphics[width=8.4cm]{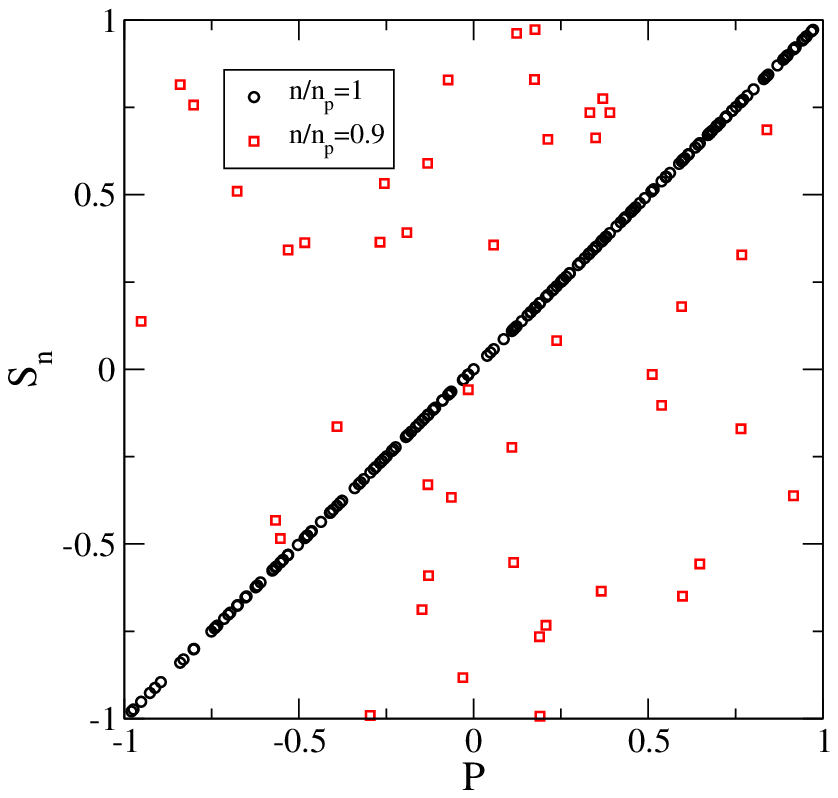} \hspace{0.5cm}
\includegraphics[width=8cm]{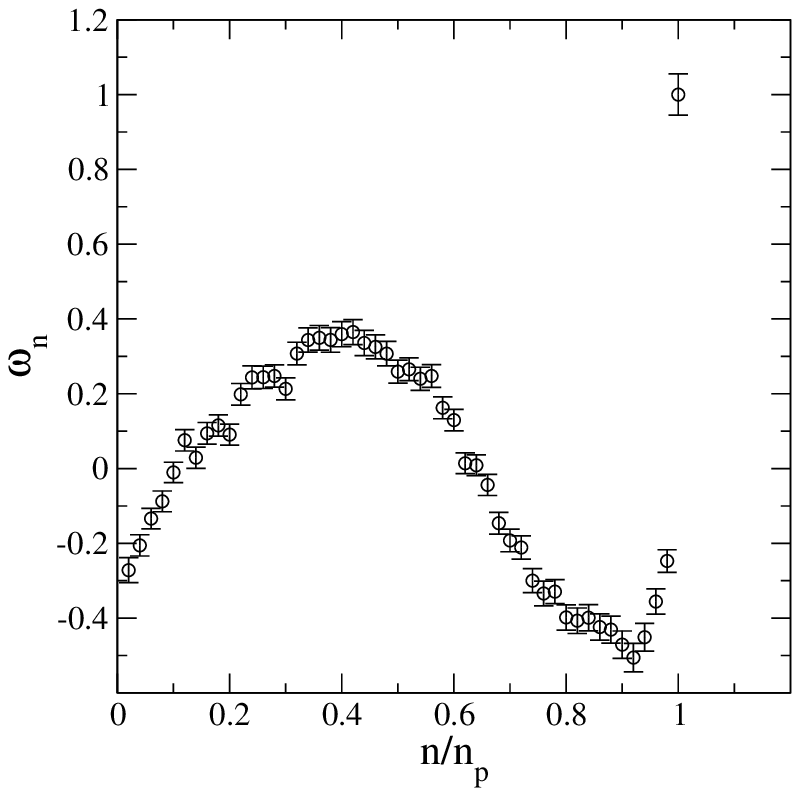}
}
\caption{ Correlation between the Polyakov loop $P(\mbx)$ and
the mode sum $\cS_n(\mbx,1)$ at each point $\mbx$ of a
single lattice configuration, $6^4$ lattice, (left).
The correlation measure $\omega_n $ (\ref{eq:k12}) as a function
of $n/n_p$, single lattice configuration, $6^4$ lattice (right).
 }
\label{fig:k2}
\end{figure*}
Using the eigenvectors $\psi_p(x)$  and eigenvalues
$\lambda_p$ of the quark operator, the Polyakov loop
$P(\mbx)$ in \refs{lat3} can be reconstructed at position $\mbx$ by
\begin{equation}
\begin{split}
\cS_n(x)& :=\frac{1}{8}\sum _{p=1}^{n}
\left(\varrho_p(x)  \lambda_p^{N_t} -\rhoz_p(x)\, \lamz_p^{N_t} \right),\\
P(\mbx)&=\cS_{n_p} (x_0,\mbx),\quad \forall x_0 
\end{split}
\label{eq:k11}
\end{equation}
where $n_p =2 N_t N_s^3$ is the total number of eigenmodes. It was already
observed earlier that the mode sum in (\ref{eq:k11}) is dominated
by the high end of the Dirac spectrum. Restricting the mode sum
to a smaller upper limit $n \ll n_p$, we do not expect that
$P(\mbx)$ and $\cS_n(x)$ are correlated locally.
In order to explore which value of $n$ must be used to obtain
a satisfactory agreement, we chose $n= 0.9 \, n_p$ and produced the
scatter plot in Figure~\ref{fig:k2}, left panel.
For $n = n_p$, we observe a perfect correlation (which served as a
benchmark test for our numerical approach). Already for
$n$ as high as $0.9 \, n_p$, this correlation has disappeared.
A similar result holds for the expectation value of
the Polyakov loop \cite{Gattringer:2006ci,Bruckmann:2006kx,Synatschke:2007bz}.

In order to quantify this correlation, we introduce
\eqnl{
\omega_n = \frac{\Bigl\langle P(\mbx) \, \cS_n (x_0,\mbx)
\Bigr\rangle _{x}}{\sqrt{
 \Bigl\langle P^2(\mbx) \Bigr\rangle _{x} \,
\Bigl\langle \cS_n^2 (x_0,\mbx) \Bigr\rangle _{x} } } . }{eq:k12}
The average extends over the space-time index, and only contributions
from a single lattice configuration are taken into account.
If $P$ and $\cS_n$ are completely uncorrelated, we find (in the
confining phase)
\eqnn{\frac{
\langle P \rangle}{\sqrt{ \langle P^2 \rangle } }\approx
 0  ,\quad\frac{
\langle \cS_n \rangle}{ \sqrt{ \langle \cS^2_n  \rangle }}
\approx  0 ,\quad
\omega_n  \approx  0  . }
By contrast, if both quantities are perfectly correlated, i.e.,
$P \propto \cS_n $, we obtain $\omega _n \approx 1$.
Figure~\ref{fig:k2}, right panel, shows $\omega_n $ as a function
of $n/n_p$, which is the fraction of the spectrum which was
considered for the mode sum \refs{eq:k11}.
Although the correlation is perfect for $n/n_p \to 1$, a decent
correlation is only achieved if almost all of the spectrum is taken
into account.

\begin{figure}[t]
\centerline{
\includegraphics[width=8cm]{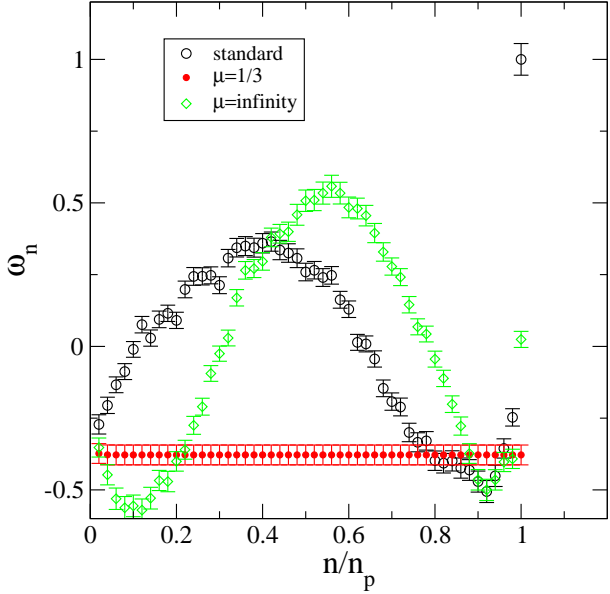}
}
\caption{
The correlation measure $\omega _n $ (\ref{eq:k12}) as a function
of $n/n_p$ for different types of mode sums,
single lattice configuration, $6^4$ lattice (right).
 }
\label{fig:k3}
\end{figure}
Let us complete this subsection by replacing the mode sum
$\cS_n$ in (\ref{eq:k11}) by the IR weighted sum
\eqnl{
\cG_n(x) :=\frac{1}{8}
\sum _{p=1}^{n} \left(\varrho_p(x) e^{- \lambda ^2_p/\mu^2 }
-\rhoz_p(x)e^{- \lamz^2_p /\mu^2 }\right)
  .}{eq:k15}
In complete analogy to (\ref{eq:k12}), we may define $\omega_n^\cG$
which quantifies the correlation between the Polyakov loop at each
point in space and $\cG_n(x_0,\mbx)$.
Figure~\ref{fig:k3} shows our findings for $\mu =1/3$ and $\mu 
\rightarrow \infty$ in comparison with $\omega_n$ in \refs{eq:k12}.
The $\mu =1/3$ graph hardly shows a dependence on $n/n_p$
simply because the mode sum is already efficiently damped by
the exponential factor. In the intermediate range $0.4 <n/n_p < 0.7$,
the mode sum utilizing $\mu \rightarrow \infty$ has a sizable correlation
with the Polyakov loop.
\subsection{ The static potential }
\begin{figure*}[t]
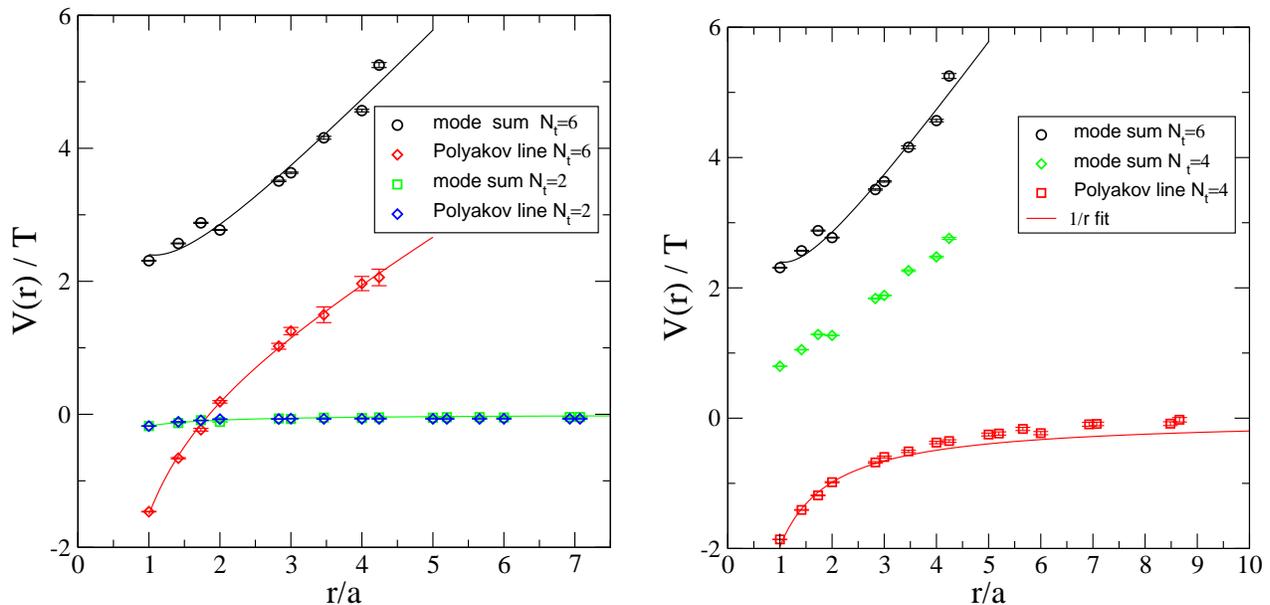

\centerline{
\includegraphics[width=8cm]{pot_low_high.eps}  \hspace{.5cm}
\includegraphics[width=8cm]{pot_crit.eps}
}
\caption{ The static quark potential extracted from the truncated
mode sum (\ref{eq:k21}) and from the Polyakov loop correlator
(\ref{eq:k20}) in the confinement phase ($N_t=6$) and in
deconfinement phase $(N_t=2)$ (left panel). The potentials for
$T \approx T_c$, $(N_t=4)$  (right panel).
}
\label{fig:k4}
\end{figure*}
On the other hand, we do expect that the expectation value
for the Polyakov loop correlator
\eqnl{
C(r)  = \Bigl\langle P(\mbx) \, P(\mbx+r \mbe_3)
\Bigr\rangle }{eq:k20}
is dominated by the IR regime of Yang-Mills theory at least for sufficiently
large separations $r$. We therefore define the mode sum approximation
to $C(\mbx,r)$ by
\eqnl{
C^{\,\cG}_n (r)=   \Bigl\langle \cG_n
(x_0,\mbx)  \, \cG_n (x_0,\mbx+r\mbe_3)
\Bigr\rangle   . }{eq:k21}
Note that the expectation value on the right hand side of the latter 
equation does not depend on the particular choice for $x_0$ due 
to translational invariance. 
From the above correlation functions, the static quark potential
$V(r)$ and its mode sum approximation $V^{\,\cG}(r)$ is obtained from
\eqnn{
V(r)/T = -  \ln C(r)  , \quad
V^{\,\cG}(r)/T = -  \ln C^{\,\cG}_n (r)  ,
}
where $T$ is the temperature.
In this first numerical study, we have chosen the exponential mode sum
with $\mu^2 = 0.1$ and truncated the mode sum by setting $n = 50$.
Thus, the lowest $50$ eigenvalues contributed.

We also investigated the deconfinement phase transition which takes place
if the temperature exceeds the critical value $T_c \approx
0.69(2) \, \sqrt{\sigma } $~\cite{Fingberg:1992ju}. Temperature is adjusted
by varying the temporal extent $N_t$ of the $N_t\times N_s^3$ lattice:
\eqnn{
\frac{T}{\sqrt{\sigma}} = \frac{1}{ N_t \, \sqrt{\sigma a^2} } \; .
}
Our simulations parameters are summarized in
Table~\ref{table:SimulationParameters}.

\begin{table}
\caption{\label{table:SimulationParameters} simulation parameters}
\begin{ruledtabular}
\begin{tabular}{ccccc}
$\beta $ &  $\sigma a^2 $ & lattice  & $T/T_c$ & configurations \\ \hline
$1.35$   &  $0.1244(7)$ & $12^3 6$ & $0.7 $ & $8658$ \\
$1.35$   &  $0.1244(7)$ & $12^3 4$ & $1.0 $ & $12000$ \\
$1.35$   &  $0.1244(7)$ & $12^3 2$ & $2.1 $ & $12000$
\end{tabular}
\end{ruledtabular}
\end{table}

Figure~\ref{fig:k4}, left panel,
shows the potentials $V^{\cG}(r)$ in comparison with
the full static potential $V(r)$ in the confinement phase for $N_t=6$.
Most striking is that the potential $V^{\cG}(r)$ linearly rises at
large distances $r$ and therefore shows confining behavior.
At small distances $r$, the potential $V^{\cG}(r)$ is flat
and does not show any sign of Coulomb law. This was not expected since
the $1/r$ part arises from the exchange of gluons and belongs to the
realm of the UV regime. The rather flat behavior implies a constant
correlation at short distances and points toward a rather smooth
field $\cG_n(x)$.

Also shown in Figure~\ref{fig:k4} are both potentials, i.e.,
$V^{\cG}(r)$  and $V(r)$, in high temperature regime. At temperatures
$T \approx 2 T_c$, both potential are essentially flat, and, in particular
$V^{\cG}(r)$, has lost any signal of the linear rise.

The situation is less clear for $T \approx T_c$ in Figure~\ref{fig:k4},
right panel. While the Polyakov loop correlator can be, to a good
extent, fitted by a $1/r$ Coulomb law, the mode sum still shows
a significant linear rise. The interesting question is whether
in the case of the mode sum
the shift of the critical temperature to higher values
is nonvanishing in the continuum limit. To answer this question,
much more time consuming simulations using higher values of $\beta $
and therefore larger lattices are necessary. It would also be interesting
to study the critical temperature signalled by $V^{\cG}(r)$ for
several types of mode sums. This is left to future work.

\subsection{Visualization}

\begin{figure}[t]
\centerline{
\includegraphics[width=8cm]{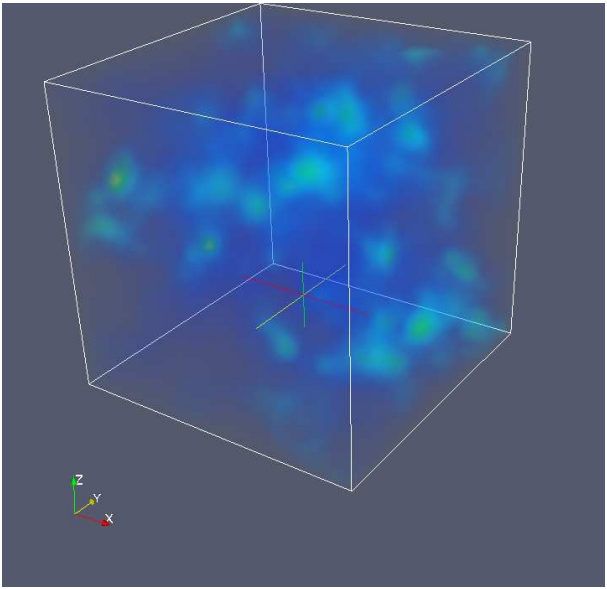}
}
\caption{ The mode sum $ \vert \cG_n(x) \vert $ in a $20^3$
spatial 
hypercube. Side length of the cube: $3.2 \,$fm.
}
\label{fig:k5}
\end{figure}
The potential $V^{\cG}(r)$ flattens for small quark antiquark 
distances $r$. This indicates that the mode sums $\cG_n(x)$ in 
(\ref{eq:k15}) are rather smooth functions of $x$. 
In order to get a first impression of the space texture of these mode sums, in
Figure~\ref{fig:k5}
we have visualized 
$$ 
\vert \cG_n(x) \vert, \quad n=25, \quad \mu = \frac{1}{3}
$$
within the spatial hypercube of a particular $20^3 \times 8 $ 
lattice configuration. The lattice configuration was generated 
with the improved action~\cite{Langfeld:2007zw} using $\beta =1.35$. 
Using $\sigma = (440 \mathrm{MeV})^2$ and $\sqrt{\sigma a^2} \approx 0.124$, 
the length of the cube is roughly $3.2 \, $fm. Thus, we 
observe a texture which is rather smooth at the length scale of $0.3 \, $fm. 
This explains the behavior of the mode sum correlator (\ref{eq:k21}) 
at short distances.

\section{Conclusions} 

It is generally accepted that the low lying modes of the quark 
operator bear witness of the spontaneous breakdown of chiral 
symmetry in QCD. Here, we have investigated to which extent the low 
lying modes also contain information on quark confinement. 

For this purpose, Gattringer's mode sum approach~\cite{Gattringer:2006ci} to
the Polyakov loop expectation value $\sum _x P(\mbx)$ has been generalized to 
reconstruct the Polyakov loop locally. We have also pointed out that 
the mode sum approach is not restricted to lattice Dirac operators, but can 
be directly formulated in the continuum formulation of Yang-Mills theory. 
The existence of these sums has been studied in some detail.
If one first sums over the center elements and afterwards over
the eigenvalues, then the spectral
sums exist \emph{for all} polynomial functions of the Dirac operator,
especially for the IR dominated mode sums of interest,
for example for $1/\cD_A^2$.
We have argued that the IR dominated mode sums equally well form an order
parameter for confinement since these sums and the Polyakov loop  share the 
same center transformation property. 

We have thoroughly investigated these mode sums by means of analytical 
calculations in the context of the Schwinger model and of $SU(2)$ gauge theory 
in the background of homogeneous field strength. 
As expected for these examples, the mode sums are proportional 
to the Polyakov loop  for each point in space. 

Subsequently, we have employed $SU(2)$ lattice gauge simulations to study the 
relation between the low lying modes of the Dirac operator and the static 
quark antiquark potential. Below the critical deconfinement temperature, 
the correlator between two IR dominated mode sums is able to describe 
a linearly rising confining potential at large distances $r$. 
In the high temperature deconfined phase, this correlator reflects 
the flat behavior of the potential at large distances, and is in very good
agreement with the correlator of two Polyakov lines. 
This clearly shows that the quark confinement mechanism is entirely encoded 
in the low lying spectrum of the Dirac operator. 
The search for confining degrees of freedom, such as vortices or monopoles, 
in the IR regime of the Dirac operator is left to future work. 

We finally point out that the preprint \cite{Bilgici:2008ui} appeared shortly
after we made
 this work available electronically. In this preprint, Bilgici and Gattringer
 also put forward the mode sum approach to the Polyakov line correlator and
 provided complementary numerical insights for the gauge group SU(3).

\begin{acknowledgements} We thank Falk Bruckmann, Christof Gattringer,
Sebastian Uhlmann and Christian Wozar for interesting discussions. 
We are grateful to Tom Heinzl for interesting comments
and to Gerald Dunne for pointing out the similarity of spectral
sums with $U(1)$-twists to the Nahm transform. The numerical calculations 
in this paper were carried out on the HPC and PlymGrid facilities 
of the University of Plymouth. 

\end{acknowledgements}
\appendix*
\section{Poisson resummation for harmonic oscillator eigenfunctions}
In this appendix we consider and resum the eigenfunctions of the squared Dirac 
operator $-\cD_A^2$ for the Schwinger model on the
torus. We measure Euclidean time in units of $\beta$ and 
lengths in units of $L$. We assume that the flux $\Phi=2\pi q$ entering the  
gauge potential
\eqnl{
A_0=-\Phi x_1+2\pi h_0\mtxt{and} A_1=2\pi h_1
}{osz0}
is positive. The \emph{periodic} eigenmodes on the cylinder $[0,1]\times \R$ 
have the form
\eqnl{
\chi_{p,\ell}(x)=e^{2\pi i \ell x_0}e^{2\pi i h_1x_1}\xi_{p}(y^1)}
{osz1}
with
$y^1=x_1+\frac{1}{q}\left(\ell-h_0\right), $
where $\ell\in\Z$ and the mode functions $\xi_p$ are eigenfunctions
of $a^\dagger a$,
\eqnl{
a^\dagger a\,\xi_p=2p\xi_p,\quad
a=(2\pi q)^{\half} y^1+(2\pi q)^{-\half}\pa_{y^1}}{osz3}
with $[a,a^\dagger]=2$.
The dependence on $\ell$ enters via the $\ell$-dependence
of $y^1$. The normalized modes are
\eqnl{
\xi_p(y)=c_p a^{\dagger p}\xi_0(y)}
{osz5}
with
\eqnl{
c_p^2=\frac{1}{2^pp!}\mtxt{and} \xi_0(x)={(2q)^{1/4}}\, 
e^{-\pi q x^2}.}{osz3b}
The normalized eigenfunctions on the \emph{torus} with boundary
conditions \refs{sch1} are superpositions of the modes on the cylinder,
\begin{equation}
\begin{split}
\psi_{p,\ell}(x)&=\sum_s e^{2\pi i sh_1} \chi_{p,\ell+sq}(x)\\
&=e^{-2\pi i qx_0x_1} e^{2\pi i(h_0y^0-\ell h_1/q)}
\sum_s f_p(y^1+s),\label{osz6}
\end{split}
\end{equation}
where we introduced the auxiliary function
\eqnl{
f_p(y^1)=e^{2\pi i q y^0y^1}\xi_p(y^1)\mtxt{with}
y^0=x_0+\frac{h_1}{q}.
 }{osz7} 
Here we can apply the Poisson resummation formula
\begin{equation}
\begin{split}
\sum_{s} f(y^1+s)&=\sum_m e^{-2\pi i my}\tilde f(m),\\
\tilde f(\eta)&=\int_{-\infty}^\infty dy^1\, e^{2\pi i\eta
y^1}f(y^1),\label{osz9}
\end{split}
\end{equation}
and together with $\tilde f_p(\eta)=\tilde\xi_p(\eta+qy^0)$ it
leads to
\begin{multline} 
\psi_{p,\ell}(x)=e^{-2\pi i qx_0x_1}
e^{2\pi i(h_0y^0-\ell h_1/q)}\\ \times
\sum_m e^{-2\pi i my^1}\,\tilde \xi_p(m+qy^0)\label{osz11}
\end{multline}
For the $q$ ground states the corresponding sums
are Gaussian and give rise to theta functions \cite{Sachs:1991en}.
To perform the sums for the excited states we observe 
that under a Fourier transformation the step operators \refs{osz3}
are transformed into step operators,
\eqnl{
\widetilde{(a^\dagger f)}(\eta)=i\tilde a^\dagger \tilde f(\eta),}{osz13b}
{where}
\eqnl{\tilde a^{\dagger}
=(2\pi\tilde q)^{\half}\eta-(2\pi\tilde q)^{-\half}\pa_{\eta}
}{osz13bb}
with dual 'instanton number' $\tilde q$ related to $q$ by
\eqnl{
q\tilde q=1.}{osz15}
The step operators $\tilde a,\tilde a^\dagger$ obey the same commutation 
relations as $a, a^\dagger$.
Since the ground state $\xi_0$ is transformed into the ground 
state with $\tilde q$ we conclude that
\eqnl{ 
\tilde \xi_p(\eta)=c_p\widetilde{(a^{\dagger p}\xi_0)}
=i^pc_p \tilde a^{\dagger p} \tilde\xi_0}{osz19}
with
$
\tilde \xi_0(\eta)={(2\tilde q)^{1/4}}\,e^{-\pi\tilde q\,\eta^2}$.
In calculations it is advantageous to use
Hermite polynomials $H_p$ generated by $a^\dagger$ and $\tilde a^\dagger$
\begin{equation}
\begin{split}
a^{\dagger p}\xi_0(y)&=H_p\left(\sqrt{2\pi q}\cdot y\right)\xi_0(y),\\
\tilde a^{\dagger p}\tilde \xi_0(\eta)&=H_p\left(\sqrt{2\pi\tilde q}\cdot
\eta\right)\tilde\xi_0(\eta).
\end{split}\label{osz21}
\end{equation}
In terms of these polynomials the equivalent series \refs{osz6} 
take the form
\begin{multline} 
\psi_{p,\ell}(x)=c_p\, e^{2\pi i \ell x_0}e^{2\pi i h_1x_1}\\\times
\sum_m e^{2m\pi i  q y^0}H_p\left(\sqrt{2\pi q}\,(y^1+m)\right)
\xi_0(y^1+m)
\label{osz23}
\end{multline}
and the resummed series \refs{osz11} reads
\begin{multline} 
\psi_{p,\ell}(x)=\frac{i^p c_p}{\sqrt{q}}\, e^{-2\pi i qx_0x_1}e^{2\pi i
(h_0y^0-\ell h_1/q)}\\ \times
\sum_m e^{-2m\pi i y^1}H_p\left(\sqrt{2\pi q}\,\big(y^0+\frac{m}{q}\big)\right)
\xi_0\big(y^0+\frac{m}{q}\big).
\label{osz25}
\end{multline}
In the main body of the paper we used antiperiodic eigenmodes of the 
Dirac operator. These are obtained by replacing $h_0\to h_0+1/2$ and
$\ell\to \ell-1/2$ in the results \refs{osz23} and \refs{osz25}.

\end{document}